\documentclass[twocolumn,amsmath,amssymb,aps,preprintnumbers,showpacs]{revtex4}

\usepackage{ulem}
\usepackage{graphics}
\usepackage{dcolumn}
\usepackage{bm}
\usepackage{amsmath}
\usepackage{amssymb}
\usepackage{epsfig}
\usepackage{upgreek}

\newif\iffig
\newif\ifgraph

\figtrue                     %

\graphtrue                   %

\begin{document}


\title{Improving superconducting resonators in magnetic fields by reduced field-focussing and engineered flux screening}

\author{D.~Bothner\footnote{Present address: Kavli Institute of NanoScience, Delft University of Technology, PO Box 5046, 2600 GA Delft, The Netherlands}}\email{daniel.bothner@uni-tuebingen.de}
\author{D.~Wiedmaier}
\author{B.~Ferdinand}
\author{R.~Kleiner}
\author{D.~Koelle}
\affiliation{Physikalisches Institut -- Experimentalphysik II and Center for Collective Quantum Phenomena in LISA$^+$, Universit\"{a}t T\"{u}bingen, Auf der Morgenstelle 14, 72076 T\"{u}bingen, Germany}

\date{\today}

\begin{abstract}

We experimentally investigate superconducting coplanar waveguide resonators in external magnetic fields and present two strategies to reduce field-induced dissipation channels and resonance frequency shifts.
One of our approaches is to significantly reduce the superconducting ground-plane areas, which leads to reduced magnetic field-focussing and thus to lower effective magnetic fields inside the waveguide cavity.
By this measure, the field-induced losses can be reduced by more than one order of magnitude in mT out-of-plane magnetic fields.
When these resonators are additionally coupled inductively instead of capacitively to the microwave feedlines, an intrinsic closed superconducting loop is effectively shielding the heart of the resonator from magnetic fields by means of flux conservation.
In total, we achieve a reduction of the field-induced resonance frequency shift by up to two orders of magnitude.
We combine systematic parameter variations on the experimental side with numerical magnetic field calculations to explain the effects of our approaches and to support our conclusions.
The presented results are relevant for all areas, where high-performance superconducting resonators need to be operated in magnetic fields, e.g. for quantum hybrid devices with superconducting circuits or electron spin resonance detectors based on coplanar waveguide cavities.

\end{abstract}


\maketitle

\section{Introduction}

During the last decade superconducting coplanar waveguide structures operated in magnetic fields have become of growing importance in different areas of research.
Coplanar microwave resonators are essential parts in hybrid quantum systems consisting of superconducting quantum circuits and natural spin systems, targeting advanced quantum information technology issues as long-lived quantum memories and microwave-to-optical frequency transduction of quantum states \cite{Rabl06, Petrosyan08, Imamoglu09, Hafezi12, Xiang13, OBrien14}.
Amongst the natural spin systems there are spin ensembles in solid state systems \cite{Chiorescu10, Kubo10, Schuster10a, Blencowe10, Kubo11, Amsuess11, Probst13, Grezes14}, ultracold atomic clouds \cite{Verdu09, Henschel10, Bernon13} as well as molecular magnets \cite{Jenkins13} or trapped single electrons \cite{Schuster10, Bushev11, Daniilidis13}, all requiring the application of external magnetic fields for being manipulated, controlled or trapped close to the resonator.
Other research fields recently using coplanar microwave structures in magnetic fields are broadband electron spin resonance experiments \cite{Bushev11a, Clauss13, Malissa13, Ranjan13, Wiemann15} and material characterization studies by means of microwave spectroscopy \cite{Scheffler13, Hafner14}.
For all these research fields, low microwave losses in the waveguide structures and a stable device operation in external magnetic fields are desired or even mandatory.
Without magnetic fields, low losses can best be achieved by using purely superconducting materials for the waveguide devices. 
In magnetic fields, however, complications arise due to Meissner currents and Abrikosov vortices, both changing the complex conductivity of the superconductor \cite{Gittleman68, Brandt91, Coffey91, Pompeo08}.
The sensitivity of superconductors to magnetic fields leads to field-dependent resonance frequency shifts of the resonators and to significantly increased losses due to an additional vortex resistivity \cite{Song09a, Bothner11, Andrews12}.
Hence, many experiments regarding the magnetic field induced property changes of superconducting microwave resonators have been carried out during the last years and different approaches to reduce or minimize them have been investigated \cite{Healey08, Song09, Groll10, deGraaf12, Bothner12, Bothner12a, Ghirri15, Samkharadze15}.
Here, we demonstrate that the performance of superconducting coplanar waveguide resonators in magnetic fields with respect to field-induced losses and frequency shifts can be significantly enhanced by a reduction of the superconducting ground-plane size and by using inductive instead of capacitive coupling.
By numerical simulations of the magnetic field distributions, we demonstrate that the reported improvements can be attributed to a significant reduction of field-focussing effects and by flux screening due to naturally built-in closed superconducting loops, when using inductive instead of capacitive coupling elements.
In order to avoid modifying the resonance mode shapes, the resonance frequencies and the characteristic impedance of the device by the superconducting ground plane reduction and in order to suppress the appearance of massive parasitic resonances (see Supplementary Information \cite{Supp16}), we replace the removed superconducting parts by normal-conducting metal on the chip and perform a careful analysis of the additional losses induced by this measure.
We focus on magnetic fields applied perpendicular to the superconducting film plane here, as in this configuration the fields have the highest impact to the resonator properties but cannot be avoided for many experiments, where out-of-plane fields are needed e.g. to trap ultracold atoms \cite{Bernon13, Bothner13} or where large in-plane fields in the $100\,$mT to Tesla range cannot be perfectly aligned to the chip surface, thus leading to out-of-plane components in the mT range.
Regarding hybrid quantum systems with magnetically trapped ultracold atoms coupled to a superconducting waveguide resonator, our approach inherently also satisfies the requirement for small superconducting areas on the microchip, needed to avoid significant trapping field distortions close to the chip surface \cite{Bernon13, Bothner13, Cano08}.
We emphasize that, although we added normal-conducting device elements (in order to keep the overall microwave device geometry unmodified between different resonator types), and although we present also an analysis regarding their contribution to the overall losses, the main results of this work are independent of such normal-conducting elements.
Thus, our results can be used to optimize similar devices for operation in magnetic fields without adding new loss channels in future approaches.

\section{Sample Fabrication and Characterization}

\begin{figure}[h]
\centering {\includegraphics[scale=0.55]{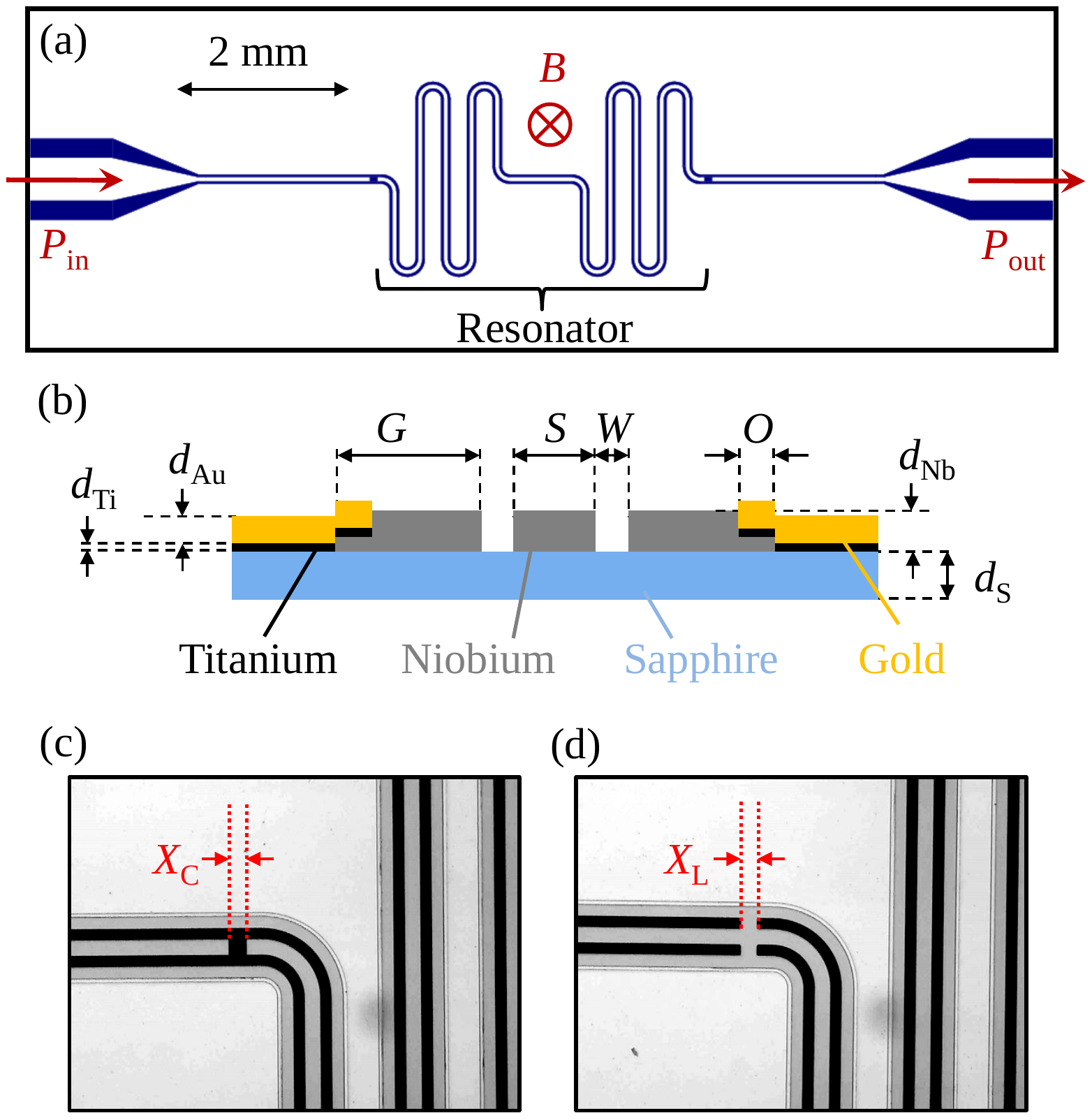}}
\caption{(Color online) (a) Layout of the $\lambda/2$-resonators used in this work. The chip size is $12\times 4\,$mm$^2$, bright parts indicate metallization, dark parts the substrate. The particular layout shown is for capacitive coupling and a center conductor width $S=50\,\mu$m. (b) Cross-sectional sketch of the normal-conducting/superconducting layer structure and the geometrical coplanar waveguide parameters. For details see text. (c), (d) Optical images of a capacitively (inductively) coupled resonator fabricated by a combination of niobium, titanium and gold. The images show a part of the resonators with $S=50\,\mu$m around the left coupling capacitor (inductor).}
\label{fig:Figure1}
\end{figure}

We have fabricated and investigated 20 different half wavelength ($\lambda/2$) resonators, all based on the same general layout, which is shown in Fig.~\ref{fig:Figure1}~(a).
On a $12\times 4\,$mm$^2$ chip, there are contact pads and straight feedlines on both sides of the meander-type resonator to send in a frequency variable microwave signal with power $P_\mathrm{in}$ on one side and to read out the transmitted power $P_\mathrm{out}$ on the other.
All resonators in our study are fabricated on r-cut sapphire wafers with a sapphire thickness of $d_\mathrm{S}=330\,\mu$m.
The resonance frequency of the fundamental mode of all resonators is designed to be $f_0=c\cdot(2\sqrt{\epsilon_\textrm{eff}}l)^{-1}=3.3\,$GHz with the effective dielectric constant $\epsilon_\textrm{eff}=5.5$, the vacuum speed of light $c$ and the length of the resonator $l=\lambda/2=19.38\,$mm.
The 20 resonators are divided into two groups of ten resonators.
The ten resonators of the first group are coupled capacitively to the feedlines, whilst the ten resonators of the second group are coupled inductively, as proposed for the realization of integrated resonator-traps in hybrid systems with ultracold atomic clouds \cite{Bothner13}.
Capacitive coupling (CC) means that there is a break in the center conductor between feedline and resonator as shown in Fig.~\ref{fig:Figure1}~(a) and (c), whereas inductive coupling (IC) is achieved by directly shortening center conductor and ground-planes as shown in Fig.~\ref{fig:Figure1}~(d).
Strictly speaking, our resonators are capacitively series-coupled or inductively shunt-coupled, and there exist also complementary coupling elements \cite{Bosman15}.
But for the sake of brevity, we will omit "series" and "shunt" in the following.
The strength of the feedline-resonator-coupling is given by the capacitance of the breaks and the inductance of the shorts, respectively, cf. Refs.~\cite{Goeppl08, Bothner13}.
In our work, all resonators are undercoupled with external quality factors $Q_\mathrm{ext}>10^5$, while the total (loaded) quality factor, measured at $4.2\,$K, is on the order of $Q\sim10^4$.
Each of the two groups (CC and IC) are further divided into two sub-sets of five resonators each.
One of the sub-sets consists of five resonators patterned from a single layer of niobium with a thickness $d_\mathrm{Nb}=500\,$nm (resonators with "extended" Nb ground-planes).
The other sub-set consists of five resonators, of which a significant part of the niobium ground-planes was removed and substituted by a $d_\mathrm{Au}=400\,$nm thick gold layer on a $d_\mathrm{Ti}=4\,$nm thick titanium sticking layer.
The center conductor and a $G=50\,\mu$m wide strip of ground-plane on each side of the center conductor still consist of niobium in the Nb/Ti/Au resonators, but the rest of the ground-plane extending to the edges of the chip is made of titanium and gold (resonators with ``narrow'' Nb ground-planes).
Between the $50\,\mu$m wide niobium ground parts and the Ti/Au parts, there is a $O=10\,\mu$m wide overlap region to ensure a good electrical contact between the layers, cf. Fig.~\ref{fig:Figure1}~(b).
Figures~\ref{fig:Figure1}~(c) and (d) show optical images of Nb/Ti/Au resonators in the vicinity of (c) a coupling gap with width $X_\textrm{C}$ and of (d) a coupling short with width $X_\textrm{L}$.
For an illustration of all relevant geometrical parameters, see Figs.~\ref{fig:Figure1}~(b)-(d).
As abbreviation, we use the letter \textbf{N} for pure niobium resonators and \textbf{A} for the resonators with reduced superconducting ground-plane areas.
The five different resonators within each of the four sub-sets (CC-\textbf{N}, CC-\textbf{A}, IC-\textbf{N} and IC-\textbf{A}) differ in their geometrical parameters of center conductor width $S=50, 40, 30, 20, 10\,\mu$m, gap width $W=25, 20, 16, 11, 6\,\mu$m and coupling element size $X_\textrm{C/L}=30, 24, 18, 12, 6\,\mu$m, respectively.
Here, all five combinations of $S$ and $W$correspond to a characteristic impedance $Z_0\approx50\,\Omega$.
This geometry variation of $S$ and $W$ allows for a systematic investigation of the influence of the normal-conducting parts on the resonator properties such as quality factor and resonance frequency in magnetic fields.
The full and unambiguous nomenclature for a single resonator throughout this paper is given by coupling\_type-\textbf{material}-center\_conductor\_width, e.g., CC-\textbf{A}-30 for a capacitively coupled resonator with Ti/Au ground-plane extensions and geometrical parameters $S=30\,\mu$m, $W=16\,\mu$m and $X_\mathrm{C}=18\,\mu$m.
The fabrication of the resonators started with a 2 inch r-cut sapphire wafer, on which we deposited the niobium layer by dc magnetron sputtering.
Then, we patterned the resonators by means of optical lithography and reactive ion etching (SF$_6$).
Resonators consisting only of niobium were finished at this point but resonators with reduced superconducting ground-planes needed some more steps.
To add the Ti/Au ground-planes, we first did another optical lithography step on the chips with narrow niobium grounds, hereby covering the superconducting parts and the gaps between center conductor and ground with photoresist, except for the $10\,\mu$m overlap region at the outer edge of the ground conductors.
We cleaned the unprotected wafer surface and at the same time removed about $200\,$nm of niobium (including a possible oxide-layer on top of the niobium) in the overlap region by a combination of Ar ion milling and SF$_6$ reactive ion etching.
Afterwards, we in-situ deposited the titanium adhesion layer by electron beam evaporation and the gold layer by dc magnetron sputtering.
Finally, we removed photoresist, Au and Ti on the superconducting parts by means of a lift-off process in warm acetone and cut the wafer into single $12\times4\,$mm$^2$ chips.
For the resonator characterization, the chips were mounted into individual, but identical brass boxes.
We used silver paste to connect the ground-planes at their edges to the box walls and to connect the on-chip feedlines to the center pin of SMA (Sub-Miniature A) connectors going through the box walls.
In order to investigate the resonators in an external magnetic field perpendicular to the chip, the boxes were mounted into the center region of a pair of Helmholtz coils with $-4\,$mT$\,\leq B\leq4\,$mT.
The box and coil configuration was built into a dip-stick with two coaxial lines and dc cables for the coil current.
All measurements presented here have been taken in liquid helium, i.e., at a temperature $T=4.2\,$K, and after zero-field cooling.
The microwave sent to the resonator was generated by a microwave signal generator and the transmitted power was measured with a spectrum analyzer.
The resonance frequency $f_0(B)$ and the quality factor $Q(B)=f_0(B)/\Delta f(B)$ were determined by fitting the measured resonance curves with a Lorentzian line and by extracting center frequency $f_0(B)$ and full width at half maximum $\Delta f(B)$ from these fits.
The measurements have been taken with applied powers between $P_\textrm{app}=-20\,$dBm and $P_\textrm{app}=0\,$dBm, leading to an approximately 5-10 dB smaller $P_\textrm{in}$ at the chip due to cable and connector losses.
While most of the characterizations have been perfomed with $P_\textrm{app}=-20\,$dBm, for some measurements it has been necessary to increase the power in order to have a reliably fittable resonance curve for all magnetic field values.
This was not only due to a field induced suppression of the resonance curve, but also due to different coupling strengths to the feedlines as well as to a varying quality of the contacts between SMA connector and lauch pads on the chip.
Hence, we have also checked the results regarding a possible power dependence but did not find any significant change with power in the investigated range.

\section{Quality and Loss Factors}
\label{sec:Losses}

\begin{figure}[h]
\centering {\includegraphics[scale=0.62]{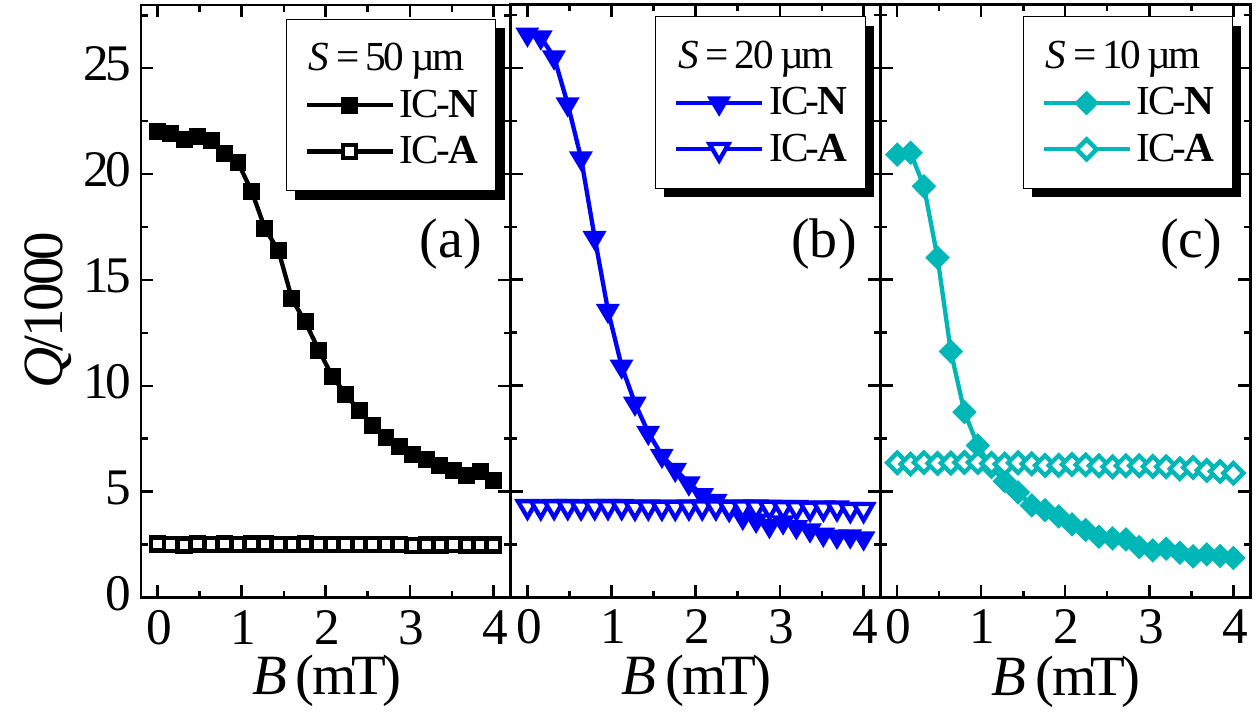}}
\caption{(Color online) Comparison of absolute quality factors $Q$ vs magnetic field $B$ between \textbf{N}- and \textbf{A}-resonators for three different geometrical parameters. All resonators are inductively coupled and the center conductor width is (a) $S=50\,\mu$m, (b) $S=20\,\mu$m, and (c) $S=10\,\mu$m.}
\label{fig:Figure2}
\end{figure}

In this section, we present data regarding the absolute quality factors $Q$ with and without magnetic field and compare $Q$ and the field-induced loss factor $1/Q_\textrm{v}$ between \textbf{N}- and \textbf{A}-resonators for different geometrical waveguide parameters.
Figure~\ref{fig:Figure2} shows, in direct comparison between \textbf{N}- and \textbf{A}-resonators, the loaded quality factor $Q$ vs applied magnetic field $B$ for three different IC resonator geometries.
For all three geometries, the quality factor of the \textbf{N}-resonators gets significantly reduced with increasing magnetic field, indicating additional losses induced by Abrikosov vortices \cite{Song09, Bothner11}, while $Q(B)$ of the \textbf{A}-resonators remains nearly constant.
Regarding the comparison between the \textbf{N}- and \textbf{A}-resonator with $S=50\,\mu$m, cf. Fig.~\ref{fig:Figure2}~(a), the \textbf{N}-resonator has a higher absolute $Q$ for all applied $B$ values and we find similar results for IC-40 and IC-30.
For $S=20\,\mu$m and $S=10\,\mu$m, however, the quality factors of the \textbf{A}-resonators exceed the $Q$ of the corresponding \textbf{N}-resonators above a certain value of $B$, which depends on $S$ and is smaller for $S=10\,\mu$m.
Moreover, the improvement in $Q$ is not only shifted to a smaller $B$-value for $S=10\,\mu$m with respect to $S=20\,\mu$m but also the ratio between $Q_\textrm{\textbf{A}}$ and $Q_\textrm{\textbf{N}}$ increases with decreasing $S$.
This is observed for all $S$ and can be attributed to two different mechanisms.

\begin{figure}[h]
\centering {\includegraphics[scale=0.7]{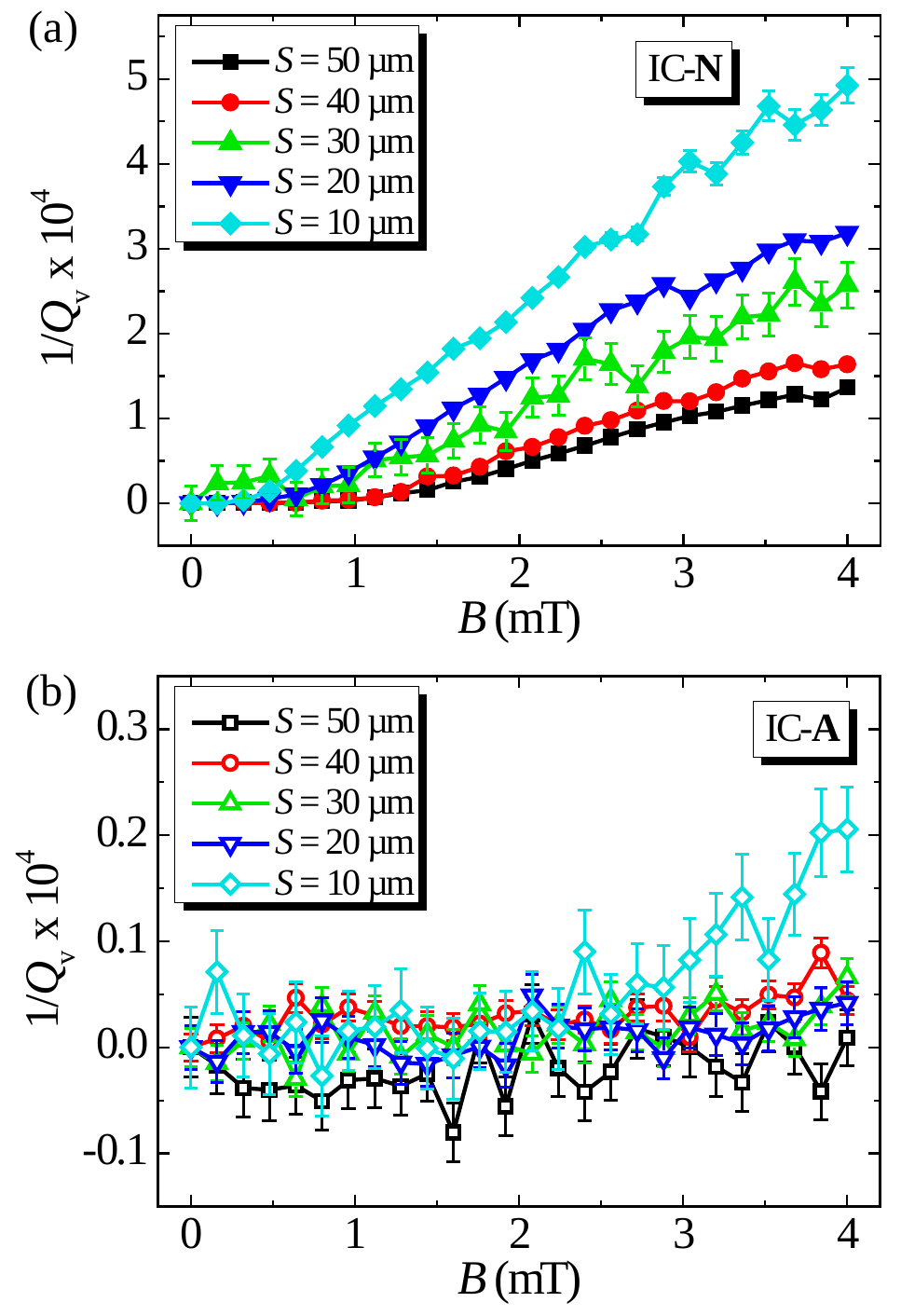}}
\caption{(Color online) Magnetic field induced loss factor $1/Q_\textrm{v}$ vs magnetic field $B$ for IC-resonators with five different geometrical parameters. (a) shows data of resonators with extended superconducting ground-planes, (b) of resonators with reduced superconducting ground-planes.}
\label{fig:Figure3}
\end{figure}

\begin{figure}[h]
\centering {\includegraphics[scale=0.6]{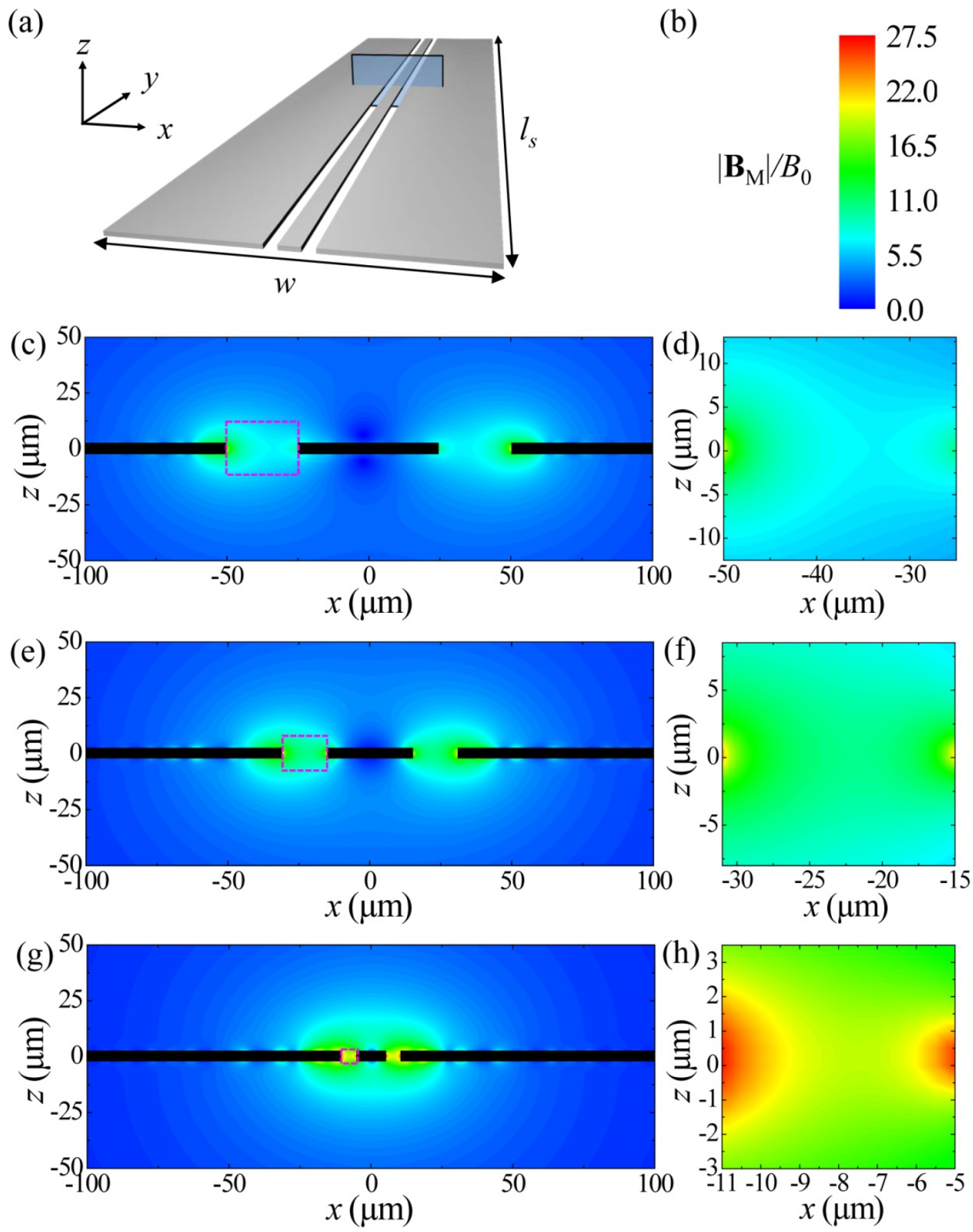}}
\caption{(Color online) \textbf{N}-type resonators: Simulation results of the static magnetic field distribution for different waveguide geometries in an external field. (a) shows a sketch of the geometry, for which the field distribution has been calculated. The transparent rectangle in the center indicates the $x$-$z$-plane, for which the magnetic field is shown in (c)-(h). The length of the structure was $l_s=3\,$mm for all simulations, the total width $w=2\,$mm, the niobium thickness $d_\textrm{Nb}=500\,$nm (from $z=0$ to $z=500\,$nm) and the London penetration depth $\lambda_\textrm{L}=120\,$nm. The externally applied field was $B_0$ in $z$-direction. (b) shows the color map used for (c)-(h). (c)-(h) show the magnitude of the calculated field $|\textbf{B}_\textrm{M}|$ normalized to the magnitude of the applied field $B_0$ for $S=50\,\mu$m, $W=25\,\mu$m [(c) and (d)], $S=30\,\mu$m, $W=16\,\mu$m [(e) and (f)] and $S=10\,\mu$m, $W=6\,\mu$m [(g) and (h)]. In (c), (e) and (g) black bars indicate the superconducting leads (thickness not to scale) and the dashed boxes show the regions in the gaps which are shown in more detail in (d), (f) and (h).}
\label{fig:Figure4}
\end{figure}

The first mechanism is an increase of magnetic-field induced losses with decreasing $S$ for the \textbf{N}-resonators.
We determine the magnetic-field induced loss factor $1/Q_\textrm{v}$ from the quality factors in magnetic field $Q(B)$ and the quality factor in zero-field $Q(B=0)$ by $1/Q_\textrm{v}=1/Q(B)-1/Q(0)$, by which all field independent loss factors are eliminated.
Figure~\ref{fig:Figure3}~(a) shows $1/Q_\textrm{v}$ for all five IC-\textbf{N} resonators vs magnetic field $B$.
These data show that with decreasing geometrical parameters the field induced loss factor gets higher.
Qualitatively and quantitatively we get almost identical results for the capacitively coupled resonators, so -- at least for \textbf{N}-resonators -- the coupling type does not have a significant influence on the field-induced loss factor.
The shown increase of field-induced losses with decreasing $S$ can be understood by the fact that with decreasing $S$, also $W$ gets smaller and the field enhancement in the gaps gets higher.
A higher field enhancement lowers the value of the applied field, at which the first vortices enter the sample.
Moreover, for a given $B$ above the critical value for vortex penetration, a stronger field enhancement also leads to an increased number of vortices inside the superconducting leads, which is connected to an increase of dissipation.
To illustrate the field enhancement in the gap and its increase with decreasing geometrical parameters, we have numerically calculated the static magnetic field $\textbf{B}_\textrm{M}$ in the Meissner state for a superconducting geometry as shown in Fig.~\ref{fig:Figure4}~(a) by solving the Maxwell and London equations with the software package 3D-MLSI \cite{Khapaev03}.
The geometry in Fig.~\ref{fig:Figure4} is a simplified version of our waveguide structures with extended niobium ground-planes.
For the simulations, we have assumed an external magnetic field of magnitude $B_0$ applied perpendicular to the plane of the superconductor with thickness $500\,$nm and a London penetration depth $\lambda_\textrm{L}=120\,$nm.
Figures~\ref{fig:Figure4}~(c)-(g) show the magnitude of the calculated field normalized to the applied field around the center conductor and in particular inside the gaps for the three parameter sets corresponding to \textbf{N}-50, \textbf{N}-30 and \textbf{N}-10.
The shown results clearly demonstrate the increasing field enhancement inside the waveguide gaps with decreasing $S$ and $W$.
Finding a direct and quantitative relation between the simulation results and the experimental data, however, is difficult due to the difference between the calculated field (Meissner state) and the vortex state responsible for the observed losses.
The field enhancement inside the gaps might also not be the only mechanism contributing to the difference in the loss factors for different $S$ and $W$.
A change of the geometrical parameters also changes the microwave current density in the center conductor and the ground-planes \cite{Clem13}, and this change can also have an influence to the field-induced losses, as in general both, the vortex distribution due to the external field and the microwave current density distribution influence the amount of dissipation \cite{Bothner12a}.

For the \textbf{A}-resonators, the field-induced losses shown in Fig.~\ref{fig:Figure3}~(b) are at least one order of magnitude smaller than for the \textbf{N}-resonators and remain close to zero for all investigated values of $B$.
Only the \textbf{A}-resonator with $S=10\,\mu$m shows a small increase of $1/Q_\textrm{v}$ for $B\gtrsim2\,$mT but the loss factor still remains a factor of $\sim20$ smaller than $1/Q_\textrm{v}$ of the corresponding \textbf{N}-resonator.
These low field-induced losses in the \textbf{A}-resonators indicate that there is no significant amount of vortices present in the superconducting leads.
Again, this can be explained in terms of the field enhancement in the gaps between center and ground conductor, which is much smaller for the \textbf{A}-resonators than for the \textbf{N}-resonators due to the reduced width of the superconducting ground.
So for the \textbf{A}-resonators with $G=50\,\mu$m wide superconducting ground conductors, the field at the center and ground conductor edges does not exceed the critical field for vortex penetration (except maybe for $S=10\,\mu$m).
We note here, that the same data for the CC-\textbf{A}-resonators show somewhat higher values for $1/Q_\textrm{v}$, nevertheless still about one order of magnitude smaller than the corresponding CC-\textbf{N}-resonators.
The small difference between IC-\textbf{A} and CC-\textbf{A} in $1/Q_\textrm{v}$ can be explained by the closed superconducting loops of the IC-resonators formed by ground-plane, center conductor and coupling shorts, which are protecting the resonator from the applied field by means of flux conservation. 
In Sec.~\ref{sec:Freq} we will discuss this point in more detail and also present field simulation results for geometries resembling those of CC-\textbf{A}- and IC-\textbf{A}-resonators.
\begin{figure}[h]
\centering {\includegraphics[scale=0.45]{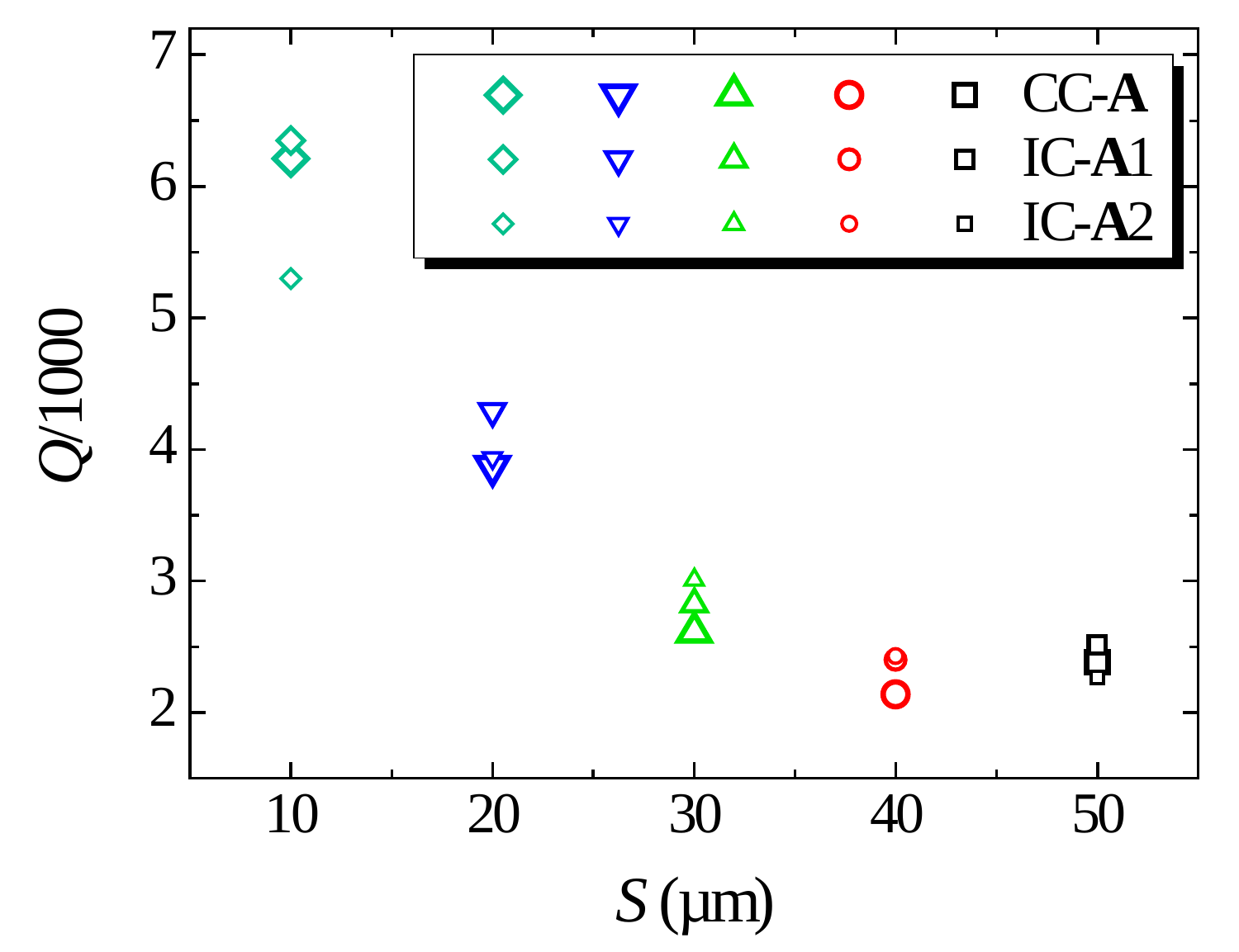}}
\caption{(Color online) Absolute zero-field quality factors of three sample-sets of \textbf{A}-resonators plotted vs center conductor width $S$. Each sample-set consists of five resonators with different geometrical parameters. One sample-set consists of capacitively coupled resonators (CC-\textbf{A}, large symbols) and two sets consist of inductively coupled resonators (IC-\textbf{A}1 and IC-\textbf{A}2, medium and small symbols).}
\label{fig:Figure5}
\end{figure}
The second mechanism, which is responsible for the observations presented in Fig.~\ref{fig:Figure2}, is related to the zero-field quality factor of the \textbf{A}-resonators depending on $S$.
Figure~\ref{fig:Figure5} shows $Q(0)$ of three sets of \textbf{A}-resonators plotted vs center conductor width $S$.
Two of the sample-sets consist of inductively coupled resonators, and one of capacitively coupled resonators.
The data reveal, that resonators with identical $S$ have a similar absolute $Q$, independent of coupling type, and the quality factors increase with decreasing $S$ from $Q\approx2500$ for $S=50\,\mu$m to $Q\approx6000$ for $S=10\,\mu$m.
The increase of $Q(0)$ with decreasing $S$ for the \textbf{A}-resonators is related to a reduced impact of the normal-conducting parts with increasing distance \textit{relative} to $S$ and $W$.
As the microwave fields in a superconducting coplanar waveguide are confined to a smaller volume around the center conductor, i.e., are decaying faster with distance, when $S$ and $W$ are smaller (cf. e.g. Ref.~\cite{Clem13}), also the microwave current density related to these fields induced at a given distance from the edge of the ground-plane gets smaller.
Hence, the contribution of a lossy material such as a normal conductor, gets reduced with increasing relative distance to center conductor and gap.
We mention the possibility here, that also the interface between the Nb and the TiAu parts contributes to the overall $Q$.
Fabricating the same resonators without etching into the Nb parts in the overlap region before depositing in-situ Ti and Au leads to zero-field quality factors, which are about one order of magnitude smaller than those shown in Fig.~\ref{fig:Figure5}.
So it is not completely clear, if the values shown in Fig.~\ref{fig:Figure5} are limited by the resistivity of Ti and Au or if a different treatment of the interface region could lead to even higher absolute $Q$s of the \textbf{A}-resonators.
The main result of this section is the observation that the magnetic-field induced losses in superconducting coplanar microwave resonators can be reduced by at least one order of magnitude when field-focussing into the waveguide gaps is suppressed by means of reducing the area of the superconducting ground-planes.
This result is independent of the presence of normal-conducting ground-plane extensions and could be fully exploited in future devices by working without normal conducting elements.
In the devices for the present experiment, however, we replaced the removed superconducting parts by normal-conducting ground-planes in order to keep the geometry unmodified, hereby adding another loss channel.
We demonstrated clearly that the new loss channel -- when full ground-planes are needed for the device -- can be controlled by adjusting the relative distance between the coplanar waveguide center and the normal-conducting parts.
For the narrowest center conductor widths used ($S = 10\,\mu$m and $S = 20\,\mu$m), we observe that the total quality factor including the normal-conductor losses can be up to a factor of three higher than that of its purely superconducting counterparts in the bulk of the investigated magnetic field range.
Additionally, in terms of total zero-field losses, even the resonator with the highest additional losses induced by the normal conductors outperforms similar, highly optimized purely normal-conducting devices by an order of magnitude \cite{Javaheri16}.

\section{Frequency Shifts and Hysteresis Effects}
\label{sec:Freq}

\begin{figure}[h]
\centering {\includegraphics[scale=0.6]{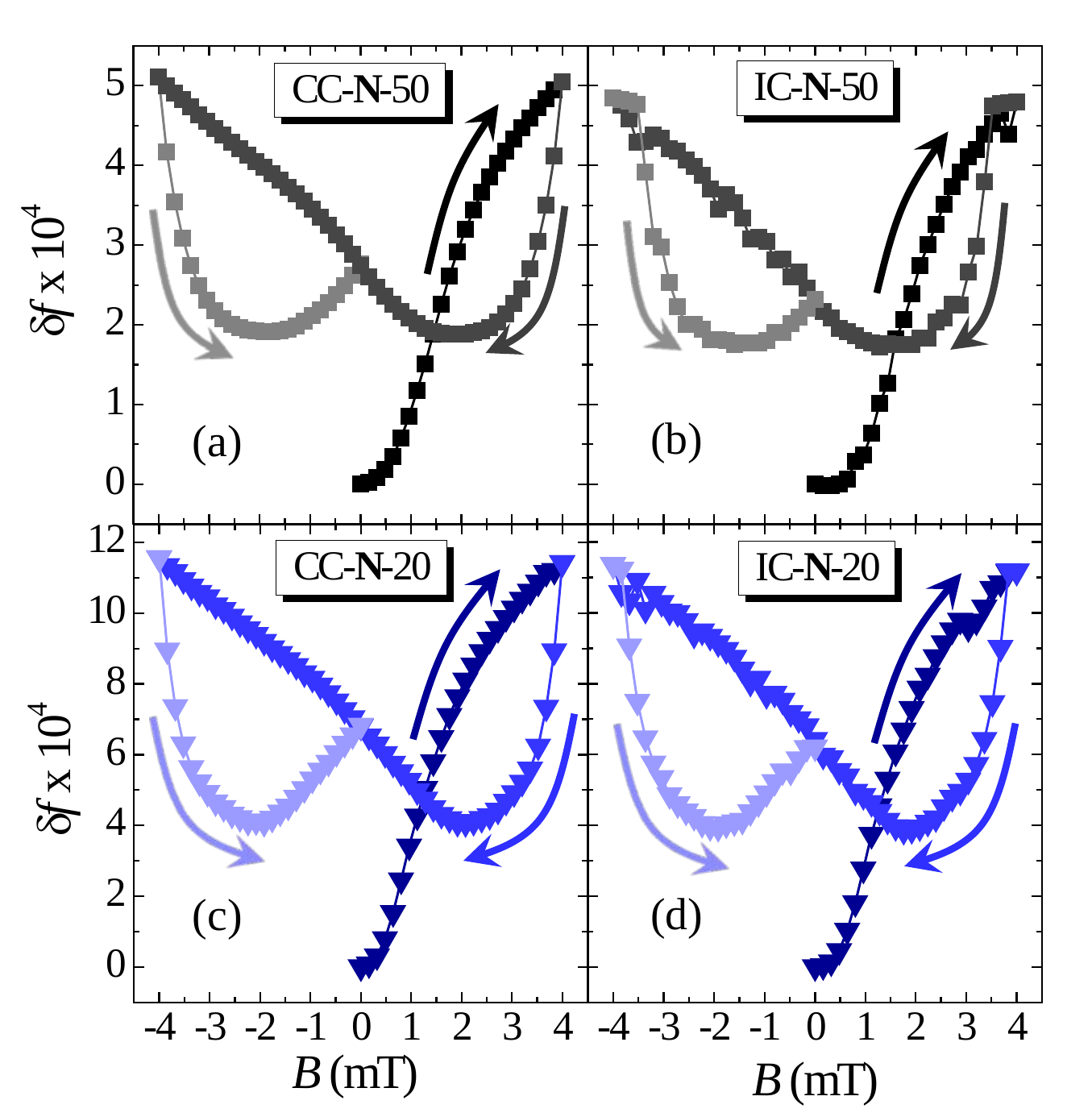}}
\caption{(Color online) Relative frequency shift $\updelta f$ (for definition see text) vs magnetic field $B$ for a magnetic field cycle $0\,$mT$\,\rightarrow4\,$mT$\,\rightarrow-4\,$mT$\,\rightarrow0\,$mT of four different \textbf{N}-type resonators. (a) shows $\updelta f$ for a CC-50, (b) for an IC-50, (c) for a CC-20, and (d) for an IC-20 resonator. Arrows indicate sweep direction.}
\label{fig:Figure6}
\end{figure}

\begin{figure}[h]
\centering {\includegraphics[scale=0.7]{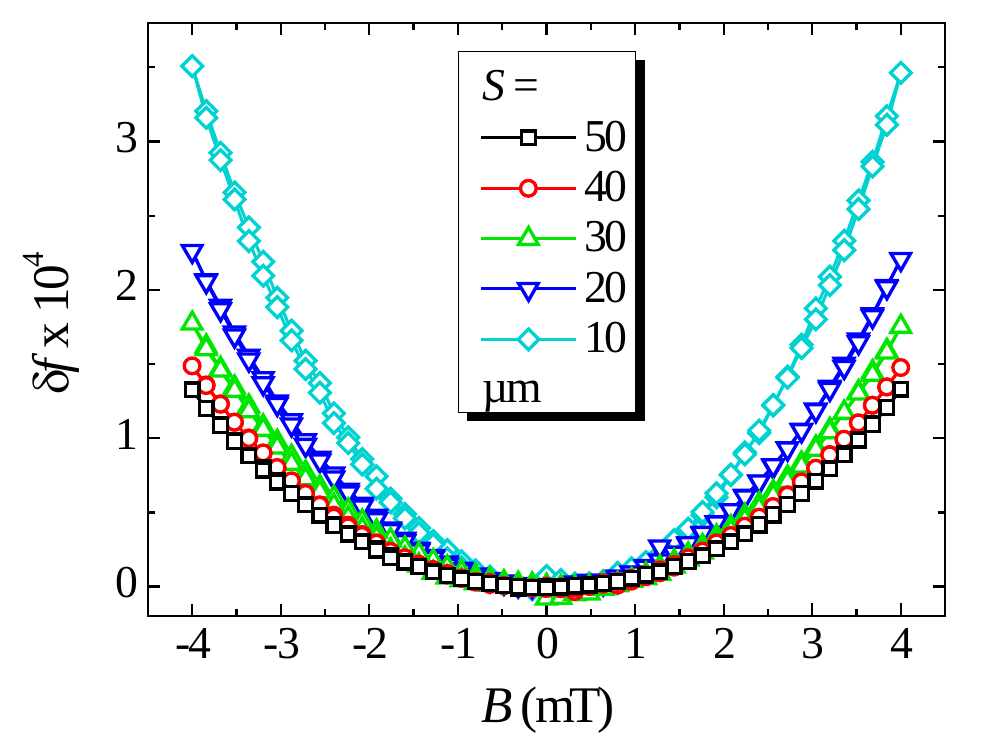}}
\caption{(Color online) Relative frequency shift $\updelta f$ vs magnetic field $B$ for a magnetic field cycle $0\,$mT$\,\rightarrow 4\,$mT$\,\rightarrow -4\,$mT$\,\rightarrow 0\,$mT of five different CC-\textbf{A}-resonators with $S=50, 40, 30, 20, 10\,\mu$m.}
\label{fig:Figure7}
\end{figure}

\begin{figure}[h]
\centering {\includegraphics[scale=0.7]{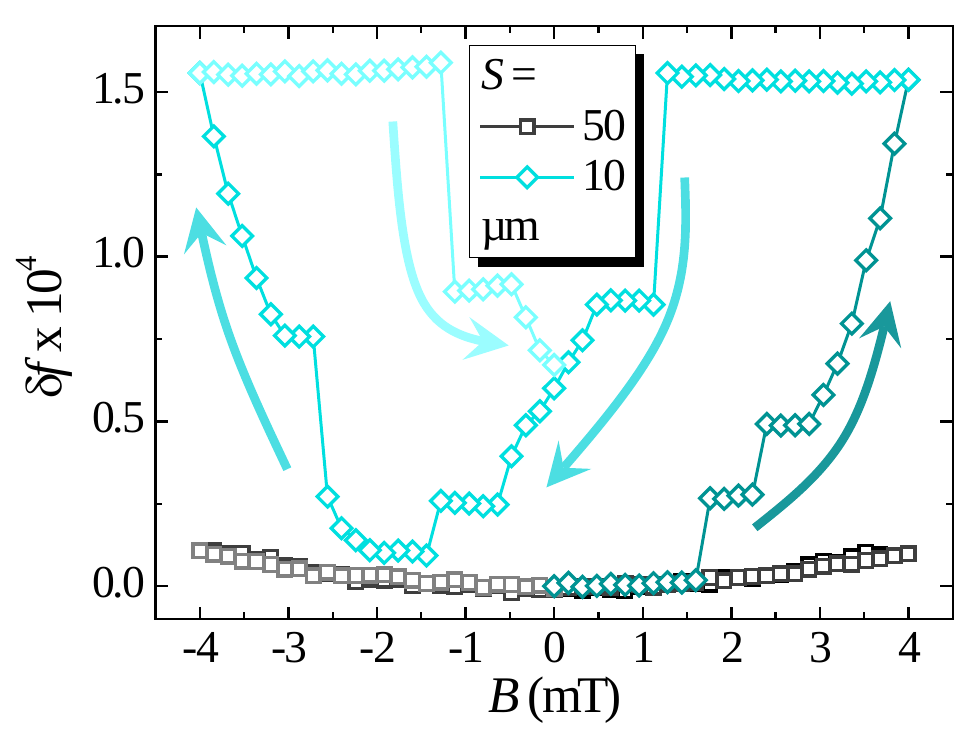}}
\caption{(Color online) Relative frequency shift $\updelta f$ vs magnetic field $B$ for a magnetic field cycle $0\,$mT$\,\rightarrow4\,$mT$\,\rightarrow-4\,$mT$\,\rightarrow0\,$mT of two different IC-\textbf{A}-resonators with $S=50\,\mu$m and $S=10\,\mu$m. Arrows indicate sweep direction.}
\label{fig:Figure8}
\end{figure}

In this section, we discuss the shift of the resonance frequency induced by the external magnetic field.
Regarding this shift, we do not only find differences between \textbf{N}- and \textbf{A}-resonators, but also between inductive and capacitive coupling. 
Moreover, we use the different hysteretic behaviours of the different resonator types to analyze the mechanisms behind the presented results in more detail.
In general, there are two magnetic field related mechanisms, which can contribute to a shift of the resonance frequency of superconducting coplanar resonators.
One of them is a change of the kinetic inductance (and thus of the total inductance) due to Meissner currents circulating in the superconductor.
This effect can be described by nonlinear London equations and leads to a non-hysteretic quadratic shift of the resonance frequency with applied magnetic field \cite{Healey08, Groll10}.
The second mechanism is related to the presence of Abrikosov vortices in the superconductor, which not only add losses but also a reactive component \cite{Gittleman68, Brandt91, Coffey91, Song09}.
If the superconductor has intrinsic defects, the Abrikosov vortices are pinned and the vortex related frequency shift is found to be hysteretic during a cycle of the static external field.
The details of the frequency shift and the hysteresis related to vortices depend on several parameters as field orientation with respect to the superconducting film, on resonator geometry as well as on the vortex distribution inside the film and the microwave current density distribution \cite{Bothner12a}.
Figure~\ref{fig:Figure6}~(a)-(d) shows the magnetic field induced relative frequency shift $\updelta f(B)=\left(f_0(0)-f_0(B)\right)/f_0(0)$ for four different \textbf{N}-resonators (two IC, two CC) during a cycle of the magnetic field from $B=0$ to $B=4\,$mT to $B=-4\,$mT and back to $B=0$.
As always in this paper, the resonators were first cooled to $T=4.2\,$K in zero magnetic field and afterwards the magnetic field was applied and swept without changing the temperature, in particular without increasing the temperature above the transition temperature again. 
For all four resonators, the frequency shift increases monotonously but nonlinearly during the virgin field upsweep.
During the downsweep from $4\,$mT to $-4\,$mT, however, the shift first reduces faster than it increased during the upsweep and after going through a minimum at still positive values of $B\approx 2\,$mT, it increases again until the sweep direction is returned.
The final upsweep from $B=-4\,$mT to $B=0$ resembles the behaviour during downsweep and if the amplitude of the sweep is not changed, the shift remains on this butterfly-like curve, also for repeated up- and downsweeps between $+4\,$mT and $-4\,$mT.
Such curves have been reported earlier for CC resonators with the same layout as used here (Fig.~\ref{fig:Figure1}~(a)) and the observed shape has been explained by a combination of a highly inhomogeneous microwave current density and the Bean model for film geometries \cite{Norris70, Brandt93}, for details see Ref.~\cite{Bothner12a}.
We observe, that the shape of the frequency shift hysteresis looks very similar for all CC-\textbf{N}-resonators, independent of $S$, the main difference between different $S$ is the magnitude of the shift, which increases with decreasing center conductor width, cf. Fig.~\ref{fig:Figure6}~(a) and (c).
This observed increase of $\updelta f$ with decreasing $S$ again can be attributed to the field enhancement in the gaps, which gets stronger with smaller $W$.
The frequency shift and the shape of the hysteresis of the IC-\textbf{N} resonators, cf. Fig.~\ref{fig:Figure6}~(b) and (d), are very similar to the ones of the CC-\textbf{N} resonators.
The absolute values of $\updelta f$ during the virgin upsweep, the minimum values during downsweep and the shape of the hystersis loop are almost identical for IC and CC.
A closer look, however, reveals two small but systematic differences between CC and IC resonators.
The first difference occurs after turning the sweep direction at $B=\pm4\,$mT.
Immediately after the turn, $\updelta f$ of the CC-resonators reduces considerably while the curve of the IC resonators shows nearly flat plateaus, where $\updelta f$ is only weakly dependent on the external field.
This effect is more pronounced for $S=50\,\mu$m, where the plateaus consist of four data points, but it is also present for $S=20\,\mu$m, where the plateaus consist of two points.
A similar effect can also be seen after starting the virgin field sweep at $B=0$.
There, $\updelta f$ of the CC-resonators increases almost immediately, while $\updelta f$ of the IC-resonators remains very small for a few data points.
The second difference between CC and IC is the smoothness of the measured curves.
For the IC resonators, the data points seem significantly more scattered (or "noisy"), which also is more pronounced for the IC-resonator with the wider center conductor.
This noisiness, however, is not due to worse Lorentzian fits or similar uncertainties during the measurement or data evaluation.
We attribute both, the small plateaus after turning the field sweep direction and the noisier curves, to the existence of closed superconducting loops in the IC resonator geometry.
The coupling shorts together with the center conductor and the ground-planes form two closed loops, and flux conservation is given for these loops until the circulating supercurrents around them exceed the critical current or vortices break into the loops by means of flux avalanches.
What is mainly relevant for field-induced losses and frequency shifts, however, is the magnetic field in the gaps between center conductor and ground-planes and at the edges of center conductor and ground conductor, respectively, where the microwave current density is peaked \cite{Healey08, Bothner12a}.
If the gaps are shielded by flux conservation, the local magnetic field at the edges of center conductor and ground-planes is not equal for IC- and CC-resonators.
The total flux and the average field in the closed loops of the IC-resonators changes stepwise each time flux breaks in, most probable when the critical current of the coupling shorts is reached.
This effect will be illuminated in more detail below, where we discuss $\updelta f$ of the IC-\textbf{A}-resonators.
First, however, we present $\updelta f$ for the five CC-\textbf{A}-resonators in Fig.~\ref{fig:Figure7}.
The data in Fig.~\ref{fig:Figure7} have been taken during a full field cycle identical to the one used for the \textbf{N}-resonators.
But the frequency shift behaves very differently here.
Independent of the center conductor width, there is no hysteresis visible, but only a reversible and quadratic shift with $B$, indicating that no vortices penetrate the superconductor.
Such a quadratic shift of the resonance frequency with the external field has been observed before and has been explained by the nonlinear Meissner effect \cite{Healey08}, i.e., by the fact that the magnetic penetration depth and hence the kinetic inductance depend quadratically on the external magnetic field.
The absence of a hysteresis is perfectly consistent with the observation that the $Q$ factors of the \textbf{A}-resonators are nearly constant in the investigated field range as discussed in Sec.~\ref{sec:Losses}.
We observe, that with decreasing gap width $W$, the field induced frequency shift increases which again can be understood by an increased field enhancement in the gaps with shrinking $W$ in combination with a geometry dependent microwave current density and kinetic inductance fraction \cite{Clem13}.
When compared with the corresponding \textbf{N}-resonators, $\updelta f$ of the \textbf{A}-resonators is smaller by a factor of $\sim 5$, demonstrating that a non-negligible fraction of $\updelta f$ in the \textbf{N}-resonators is provided by Meissner currents and the frequency shift of the \textbf{N}-resonators is not exclusively caused by vortices.
A clear separation of the shift induced by Meissner currents and the shift induced by vortices, however, is difficult for our sample geometry, because for \textbf{N}-resonators the field focusing factor depends strongly on the position along the resonator due to a position dependent ground-plane width, cf. Fig.~\ref{fig:Figure1}~(a).
Finally, we present results on $\updelta f$ for two IC-\textbf{A}-resonators ($S=50\,\mu$m and $S=10\,\mu$m) in Fig.~\ref{fig:Figure8}.
The frequency shift for the resonator with $S=50\,\mu$m shows no hysteresis and a small quadratic shift similar to its CC counterpart.
However, the magnitude of the shift is approximately one order of magnitude smaller than for the corresponding CC-resonator, cf. Fig.~\ref{fig:Figure7}.
We think that this is once more a consequence of the closed superconducting loops, which keep the total flux in the loops zero for the investigated field range.
If the interior of the loop is shielded, the local field and the Meissner currents at the sensitive inner edges of the loop, i.e., where the microwave currents are maximum, are significantly smaller (though still finite) than without the shielding effect of the loop as for the CC-resonators.
\begin{figure}[h]
\centering {\includegraphics[scale=0.7]{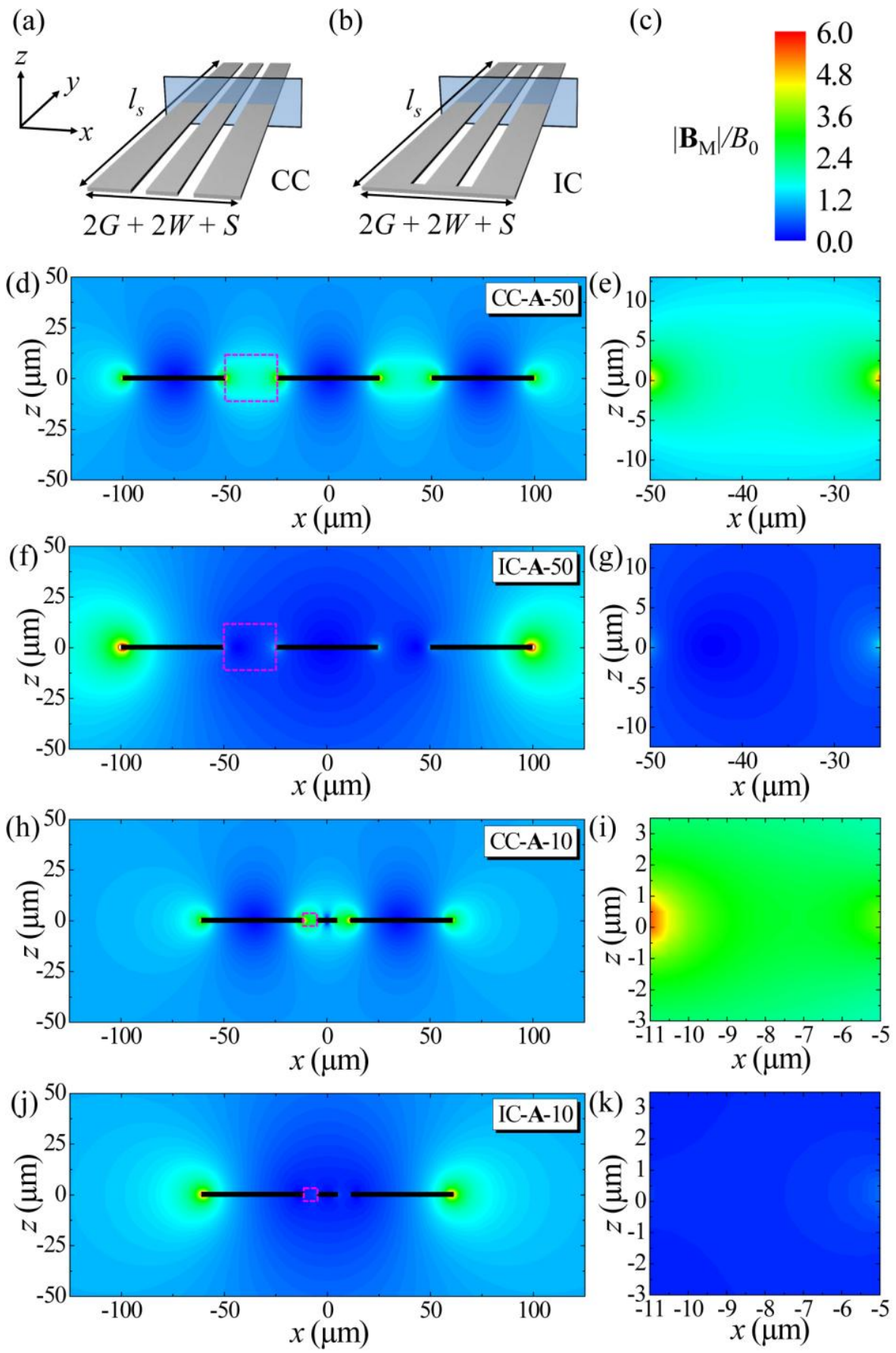}}
\caption{(Color online) \textbf{A}-type resonators: Simulation results of the static magnetic field distribution for different waveguide geometries. (a) and (b) show sketches of the two different types of geometries, for which the field distribution has been calculated. The transparent rectangle in the center indicates the $x$-$z$-plane, for which the field is shown in (d)-(k). The length of the structure was $l_s=3\,$mm for all simulations, the niobium thickness $d_\textrm{Nb}=500\,$nm (from $z=0$ to $z=500\,$nm) and the London penetration depth $\lambda_\textrm{L}=120\,$nm. The externally applied field was $B_0$ in $z$-direction. (c) shows the color map used for (d)-(k). (d)-(k) show the magnitude of the calculated field $|\textbf{B}_\textrm{M}|$ normalized to the magnitude of the applied field $B_0$ for $S=50\,\mu$m, $W=25\,\mu$m, $G=50\,\mu$m [(d)-(g)] and $S=10\,\mu$m, $W=6\,\mu$m, $G=50\,\mu$m [(h)-(k)]. (d), (e), (h) and (i) show results for the CC geometry, (f), (g), (j) and (k)  for the IC geometry. In (d), (f), (h) and (j) black bars indicate the superconducting leads (thickness not to scale) and the dashed boxes show the regions in the gaps which are shown in more detail in (e), (g), (i) and (k).}
\label{fig:Figure9}
\end{figure}
This shielding effect can also be seen in the results of numeric magnetic field calculations shown in Fig.~\ref{fig:Figure9}, where for two geometries ($S=50\,\mu$m and $S=10\,\mu$m) the static magnetic field in simplified CC-\textbf{A}- and IC-\textbf{A}-resonator structures is shown in a direct comparison between CC and IC layout.
If the field inside the gaps of the structure shown in Fig.~\ref{fig:Figure9}~(a) (CC) is calculated, we find an enhancement of the applied field as shown in (d), (e) and (h), (i).
As mentioned earlier, this field enhancement is significantly smaller than for extended ground-planes, cf. Fig.~\ref{fig:Figure4}.
If, however, the structure shown in Fig.~\ref{fig:Figure9}~(b) is used for the simulations, which resembles an IC-\textbf{A} resonator with closed superconducting loops, we find the results presented in (f), (g) and (j), (k) with comparatively small magnetic field values inside the gaps.
Instead, the field enhancement at the outer edges of the narrow Nb ground conductors gets significantly higher for the IC geometry.
There, however, the effective field has a much smaller influence on the resonator properties, because the microwave field and currents are mainly confined to the center conductor and gap region.
What furthermore supports our interpretation taking into account the closed loops, is the shape of the $\updelta f$ hysteresis for the IC-\textbf{A}-resonator with $S=10\,\mu$m, cf. the diamond shaped data points in Fig.~\ref{fig:Figure8}.
For small fields, the shift is on the same small order of magnitude than for $S=50\,\mu$m, but at $B\approx1.6\,$mT, a sudden jump occurs, after which the shift stays nearly constant until a second jump occurs at $B\approx2.3\,$mT.
During the field cycle and in particular immediately after the sweep direction turning points, some more similar plateaus and jumps occur, which are reproducible for a certain sample but which vary between different samples.
We observe that such jumps only occur for $S=10\,\mu$m and less pronounced for $S=20\,\mu$m, what can be understood by estimating the critical current of the inductance shorts, which gets smaller with decreasing $S$ due to a decreased width $X_\textrm{L}$.
With a typical critical current density $j_c\approx4\cdot10^6\,$Acm$^{-2}$ of our Nb films at $T=4.2\,$K, the critical current of a coupling short with $X_\textrm{L}=30\,\mu$m and $X_\textrm{L}=6\,\mu$m, respectively, is $I_{c50}\approx600\,$mA and $I_{c10}\approx120\,$mA ($d_\textrm{Nb}=500\,$nm).
From our simulations, we also determined the total current per mT around the loops for the different $S$ and $W$.
For $S=50\,\mu$m, we obtain $I/B_{50}\approx95\,$mA/mT and $I/B_{10}\approx45\,$mA/mT for $S=10\,\mu$m, which implies that for $S=50\,\mu$m the critical current of the coupling shorts is higher than the current they have to carry at $B=4\,$mT, whilst for $S=10\,\mu$m the critical current is reached at $B\approx2.7\,$mT.
As this consideration is not meant to be a precise analysis but more an estimation to see if our interpretation is reasonable, we do not go further into detail here.
With this latter discussion in mind, we turn back to $\updelta f$ of the IC-\textbf{N}-resonators, where we have found similar small plateaus and a kind of noise in the hysteresis curve, cf. Fig.~\ref{fig:Figure6}~(b), (d) and the corresponding discussion in the main text.
According to our interpretation of the hysteresis for IC-\textbf{A}-10, jumps and plateaus are related to exceeding the critical current of the inductance shorts and flux conservation in the loops.
For IC-\textbf{N}-resonators, the effective area $A_\textrm{eff}$ of the closed loops is much larger than for the \textbf{A}-resonators, implying that a much larger net current $I=B\cdot A_{\textrm{eff}}/L$ has to circulate around the loop for a given external field in order to ensure flux conservation.
For our geometry, this effect is even more enhanced by a slightly decreasing loop inductance $L$ with increasing ground conductor width.
An increased circulating shielding current per unit field in turn means, that the critical current of the inductance shorts is reached at a much smaller value of the applied field and the field in the loops is nearly following the external field.
Such an interpretation is also consistent with the observation that the plateaus and the noisiness are more distinct for the wider center conductor, because the smaller the center conductor is, the smaller is the field change, for which the shorts can shield the loop.
The results presented in this section demonstrate, that the magnetic field induced frequency shift in superconducting coplanar resonators can be reduced by more than one order of magnitude with inductive coupling and reduced superconducting ground-planes compared to resonators with extended superconductor ground-planes.
Depending on coupling type, geometrical parameters and superconducting ground-plane width, the resonance frequency shift varies over almost two orders of magnitude and is dominated by different mechanisms for different resonator types.
For capacitive coupling and narrow superconducting ground-planes, the shift is quadratic with the external field, caused by Meissner currents, and shows no hysteresis, independent of center conductor width.
Inductively coupled resonators with reduced superconducting ground-planes show the smallest shift due to flux conservation in the closed superconducting loops.
The shift here is also caused by Meissner currents, but is more complicated and hysteretic for narrower waveguides due to abrupt changes of the total flux inside the loops.
In resonators with extended superconducting ground-planes, Meissner currents (CC and IC) and flux conservation (IC only) contribute also to the field induced frequency shift, but the dominant mechanism are Abrikosov vortices penetrating the sample.
The vortices lead to a significant hysteresis, which is qualitatively nearly independent of coupling type and geometrical parameters.

\section{Conclusions}

In this work, we have presented a comprehensive experimental study on superconducting coplanar microwave resonators in external magnetic fields and investigated systematically the influence of reduced superconducting ground-plane areas as well as of the coupling type on field-induced changes of quality factor and resonance frequency.
We reported that by using reduced superconducting ground-plane areas the magnetic-field induced losses can be reduced by more than one order of magnitude for perpendicular magnetic fields in the mT range.
We attribute this significant loss reduction to reduced field-focussing into the waveguide gaps, which is additionally supported by the experimentally found dependence of the losses on waveguide gap width as well as by numerical simulations.
In addition to the significant reduction of losses by preventing Abrikosov vortices from entering the resonator, the sensitivity of the resonance frequency to magnetic fields can be suppressed by another order of magnitude, if instead of the usual capacitive coupling an inductive coupling approach is used.
The inductive coupling shorts hereby form closed superconducting loops, which shield the waveguide gap interior from the magnetic field by means of flux conservation as shown by numerical simulations.
In our experiments, we did not only remove parts of the superconducting ground-planes, but we replaced them with a normal conductor to ensure identical resonator modes and to suppress massive and systematic parasitic resonances appearing without proper grounding \cite{Supp16, Wenner11}.
Thus, we also characterized the additional losses induced by the normal conductors and demonstrated that, depending on the waveguide geometry, the total quality factor of a superconductor/normal-conductor hybrid resonator in a magnetic field can be significantly higher than both, the quality factor of purely superconducting resonators as well as the typical quality factor of purely normal-conducting resonators.
The use of normal-conducting components, however, might not be necessary in general and a purely superconducting resonator with small ground planes and inductive coupling would in fact profit most from carefully engineered field-focussing and by taking advantage of flux screening by means of inductive coupling.
Our findings thus pave the road towards implementing these optimization strategies into purely superconducting devices.
Finally, our strategies can be combined with earlier approaches to improve superconducting resonators in magnetic fields such as patterning microholes or slots as pinning sites for vortices into the sample.

\section{Acknowledgements}

This work has been supported by the Deutsche Forschungsgemeinschaft via the SFB/TRR 21 C2 and by the EU-FP6-COST Action MP1201.
B. F. gratefully acknowledges support by the Carl Zeiss Stiftung.


\begin{thebibliography}{26}
\expandafter\ifx\csname natexlab\endcsname\relax\def\natexlab#1{#1}\fi
\expandafter\ifx\csname bibnamefont\endcsname\relax
  \def\bibnamefont#1{#1}\fi
\expandafter\ifx\csname bibfnamefont\endcsname\relax
  \def\bibfnamefont#1{#1}\fi
\expandafter\ifx\csname citenamefont\endcsname\relax
  \def\citenamefont#1{#1}\fi
\expandafter\ifx\csname url\endcsname\relax
  \def\url#1{\texttt{#1}}\fi
\expandafter\ifx\csname urlprefix\endcsname\relax\def\urlprefix{URL }\fi
\providecommand{\bibinfo}[2]{#2}
\providecommand{\eprint}[2][]{\url{#2}}

\bibitem[{\citenamefont{Rabl \textit{et~al.}}(2006)\citenamefont{Rabl, DeMille, Doyle, Lukin, Schoelkopf, Zoller}}]{Rabl06}
  \bibinfo{author}{\bibfnamefont{P.}~\bibnamefont{Rabl}},
  \bibinfo{author}{\bibfnamefont{D.}~\bibnamefont{DeMille}},
  \bibinfo{author}{\bibfnamefont{J.~M.}~\bibnamefont{Doyle}},
  \bibinfo{author}{\bibfnamefont{M.~D.}~\bibnamefont{Lukin}},
  \bibinfo{author}{\bibfnamefont{R.~J.}~\bibnamefont{Schoelkopf}}, \bibnamefont{and}
  \bibinfo{author}{\bibfnamefont{P.}~\bibnamefont{Zoller}},
  \textit{Hybrid Quantum Processors: Molecular Ensembles as Quantum Memory for Solid State Circuits},
  \bibinfo{journal}{Phys. Rev. Lett.} \textbf{\bibinfo{volume}{97}},
  \bibinfo{pages}{033003} (\bibinfo{year}{2006})
	  
\bibitem[{\citenamefont{Petrosyan and Fleischhauer}(2008)\citenamefont{Petrosyan and Fleischhauer}}]{Petrosyan08}
  \bibinfo{author}{\bibfnamefont{D.}~\bibnamefont{Petrosyan}} \bibnamefont{and}
  \bibinfo{author}{\bibfnamefont{M.}~\bibnamefont{Fleischhauer}},
  \textit{Quantum Information Processing with Single Photons and Atomic Ensembles in Microwave Coplanar Waveguide Resonators},
  \bibinfo{journal}{Phys. Rev. Lett.} \textbf{\bibinfo{volume}{100}},
  \bibinfo{pages}{170501} (\bibinfo{year}{2008})
	  
\bibitem[{\citenamefont{Imamo\u{g}lu}(2009)\citenamefont{Imamo\u{g}lu}}]{Imamoglu09}
  \bibinfo{author}{\bibfnamefont{Ata\c{c}}~\bibnamefont{Imamo\u{g}lu}},
  \textit{Cavity QED Based on Collective Magnetic Dipole Coupling: Spin Ensembles as Hybrid Two-Level Systems},
  \bibinfo{journal}{Phys. Rev. Lett.} \textbf{\bibinfo{volume}{102}},
  \bibinfo{pages}{083602} (\bibinfo{year}{2009})
	
\bibitem[{\citenamefont{Hafezi \textit{et~al.}}(2006)\citenamefont{Hafezi, Kim, Rolston, Orozco, Lev, Taylor}}]{Hafezi12}
  \bibinfo{author}{\bibfnamefont{M.}~\bibnamefont{Hafezi}},
  \bibinfo{author}{\bibfnamefont{Z.}~\bibnamefont{Kim}},
  \bibinfo{author}{\bibfnamefont{S.~L.}~\bibnamefont{Rolston}},
  \bibinfo{author}{\bibfnamefont{L.~A.}~\bibnamefont{Orozco}},
  \bibinfo{author}{\bibfnamefont{B.~L.}~\bibnamefont{Lev}}, \bibnamefont{and}
  \bibinfo{author}{\bibfnamefont{J.~M.}~\bibnamefont{Taylor}},
  \bibinfo{Atomic interface between microwave and optical photons},
  \bibinfo{journal}{Phys. Rev. A} \textbf{\bibinfo{volume}{85}},
  \bibinfo{pages}{020302(R)} (\bibinfo{year}{2012})

\bibitem[{\citenamefont{Xiang \textit{et~al.}}(2013)\citenamefont{Xiang, Ashhab, You, Nori}}]{Xiang13}
  \bibinfo{author}{\bibfnamefont{Z.-L.}~\bibnamefont{Xiang}},
  \bibinfo{author}{\bibfnamefont{S.}~\bibnamefont{Ashhab}},
  \bibinfo{author}{\bibfnamefont{J.~Q.}~\bibnamefont{You}}, \bibnamefont{and}
  \bibinfo{author}{\bibfnamefont{F.}~\bibnamefont{Nori}},
  \textit{Hybrid quantum circuits: Superconducting circuits interacting with other quantum systems},
  \bibinfo{journal}{Rev. Mod. Phys.} \textbf{\bibinfo{volume}{85}},
  \bibinfo{pages}{623} (\bibinfo{year}{2013})

\bibitem[{\citenamefont{O'Brien \textit{et~al.}}(2014)\citenamefont{O'Brien, Lauk, Blum, Morigi, Fleischhauer}}]{OBrien14}
  \bibinfo{author}{\bibfnamefont{C.}~\bibnamefont{O'Brien}},
  \bibinfo{author}{\bibfnamefont{N.}~\bibnamefont{Lauk}},
  \bibinfo{author}{\bibfnamefont{S.}~\bibnamefont{Blum}},
  \bibinfo{author}{\bibfnamefont{G.}~\bibnamefont{Morigi}}, \bibnamefont{and}
  \bibinfo{author}{\bibfnamefont{M.}~\bibnamefont{Fleischhauer}},
  \textit{Interfacing Superconducting Qubits and Telecom Photons via a Rare-Earth-Doped Crystal},
  \bibinfo{journal}{Phys. Rev. Lett.} \textbf{\bibinfo{volume}{113}},
  \bibinfo{pages}{063603} (\bibinfo{year}{2014})
	
\bibitem[{\citenamefont{Chiorescu \textit{et~al.}}(2010)\citenamefont{Chiorescu, Groll, Bertaina, Mori, Miyashita}}]{Chiorescu10}
  \bibinfo{author}{\bibfnamefont{I.}~\bibnamefont{Chiorescu}},
  \bibinfo{author}{\bibfnamefont{N.}~\bibnamefont{Groll}},
  \bibinfo{author}{\bibfnamefont{S.}~\bibnamefont{Bertaina}},
  \bibinfo{author}{\bibfnamefont{T.}~\bibnamefont{Mori}}, \bibnamefont{and}
  \bibinfo{author}{\bibfnamefont{S.}~\bibnamefont{Miyashita}},
  \textit{Magnetic strong coupling in a spin-photon system and transition to classical regime},
  \bibinfo{journal}{Phys. Rev. B} \textbf{\bibinfo{volume}{82}},
  \bibinfo{pages}{024413} (\bibinfo{year}{2010})
	
\bibitem[{\citenamefont{Kubo \textit{et~al.}}(2010)\citenamefont{Kubo, Ong, Bertet, Vion, Jacques, Zheng, Dr\'eau, Roch, Auffeves, Jelezko, Wrachtrup, Barthe, Bergonzo, Esteve}}]{Kubo10}
  \bibinfo{author}{\bibfnamefont{Y.}~\bibnamefont{Kubo}},
  \bibinfo{author}{\bibfnamefont{F.~R.}~\bibnamefont{Ong}},
  \bibinfo{author}{\bibfnamefont{P.}~\bibnamefont{Bertet}},
  \bibinfo{author}{\bibfnamefont{D.}~\bibnamefont{Vion}},
  \bibinfo{author}{\bibfnamefont{V.}~\bibnamefont{Jacques}},
  \bibinfo{author}{\bibfnamefont{D.}~\bibnamefont{Zheng}},
  \bibinfo{author}{\bibfnamefont{A.}~\bibnamefont{Dr\'eau}},
  \bibinfo{author}{\bibfnamefont{J.-F.}~\bibnamefont{Roch}},
  \bibinfo{author}{\bibfnamefont{A.}~\bibnamefont{Auffeves}},
  \bibinfo{author}{\bibfnamefont{F.}~\bibnamefont{Jelezko}},
   \bibinfo{author}{\bibfnamefont{J.}~\bibnamefont{Wrachtrup}},
  \bibinfo{author}{\bibfnamefont{M.~F.}~\bibnamefont{Barthe}},
  \bibinfo{author}{\bibfnamefont{P.}~\bibnamefont{Bergonzo}}, \bibnamefont{and}
  \bibinfo{author}{\bibfnamefont{D.}~\bibnamefont{Esteve}},
  \textit{Strong Coupling of a Spin Ensemble to a Superconducting Resonator},
  \bibinfo{journal}{Phys. Rev. Lett.} \textbf{\bibinfo{volume}{105}},
  \bibinfo{pages}{140502} (\bibinfo{year}{2010})
 
\bibitem[{\citenamefont{Schuster \textit{et~al.}}(2010)\citenamefont{Schuster, Sears, Ginossar, DiCarlo, Frunzio, Morton, Wu, Briggs, Buckley, Awschalom, Schoelkopf}}]{Schuster10a}
  \bibinfo{author}{\bibfnamefont{D.~I.}~\bibnamefont{Schuster}},
  \bibinfo{author}{\bibfnamefont{A.~P.}~\bibnamefont{Sears}},
  \bibinfo{author}{\bibfnamefont{E.}~\bibnamefont{Ginossar}},
  \bibinfo{author}{\bibfnamefont{L.}~\bibnamefont{DiCarlo}},
  \bibinfo{author}{\bibfnamefont{L.}~\bibnamefont{Frunzio}},
  \bibinfo{author}{\bibfnamefont{J.~J.~L.}~\bibnamefont{Morton}},
  \bibinfo{author}{\bibfnamefont{H.}~\bibnamefont{Wu}},
  \bibinfo{author}{\bibfnamefont{G.~A.~D.}~\bibnamefont{Briggs}},
  \bibinfo{author}{\bibfnamefont{B.~B.}~\bibnamefont{Buckley}},
  \bibinfo{author}{\bibfnamefont{D.~D.}~\bibnamefont{Awschalom}}, \bibnamefont{and}
  \bibinfo{author}{\bibfnamefont{R.~J.}~\bibnamefont{Schoelkopf}},
  \textit{High-Cooperativity Coupling of Electron-Spin Ensembles to Superconducting Cavities},
  \bibinfo{journal}{Phys. Rev. Lett.} \textbf{\bibinfo{volume}{105}},
  \bibinfo{pages}{140501} (\bibinfo{year}{2010})
	
\bibitem[{\citenamefont{Blencowe}(2010)\citenamefont{Blencowe}}]{Blencowe10}
  \bibinfo{author}{\bibfnamefont{Miles}~\bibnamefont{Blencowe}},
  \textit{Quantum Computing: Quantum RAM},
  \bibinfo{journal}{Nature} \textbf{\bibinfo{volume}{468}},
  \bibinfo{pages}{44} (\bibinfo{year}{2010})
 
\bibitem[{\citenamefont{Kubo \textit{et~al.}}(2011)\citenamefont{Kubo, Grezes, Dewes, Umeda, Isoya, Sumiya, Morishita, Abe, Onoda, Ohshima, Jacques, Dr\'eau, Roch, Diniz, Auffeves, Vion, Esteve, Bertet}}]{Kubo11}
  \bibinfo{author}{\bibfnamefont{Y.}~\bibnamefont{Kubo}},
  \bibinfo{author}{\bibfnamefont{C.}~\bibnamefont{Grezes}},
  \bibinfo{author}{\bibfnamefont{A.}~\bibnamefont{Dewes}},
  \bibinfo{author}{\bibfnamefont{T.}~\bibnamefont{Umeda}},
  \bibinfo{author}{\bibfnamefont{J.}~\bibnamefont{Isoya}},
  \bibinfo{author}{\bibfnamefont{H.}~\bibnamefont{Sumiya}},
  \bibinfo{author}{\bibfnamefont{N.}~\bibnamefont{Morishita}},
  \bibinfo{author}{\bibfnamefont{H.}~\bibnamefont{Abe}},
  \bibinfo{author}{\bibfnamefont{S.}~\bibnamefont{Onoda}},
  \bibinfo{author}{\bibfnamefont{T.}~\bibnamefont{Ohshima}},
  \bibinfo{author}{\bibfnamefont{V.}~\bibnamefont{Jacques}},
  \bibinfo{author}{\bibfnamefont{A.}~\bibnamefont{Dr\'eau}},
  \bibinfo{author}{\bibfnamefont{J.-F.}~\bibnamefont{Roch}},
  \bibinfo{author}{\bibfnamefont{I.}~\bibnamefont{Diniz}},
  \bibinfo{author}{\bibfnamefont{A.}~\bibnamefont{Auffeves}},
  \bibinfo{author}{\bibfnamefont{D.}~\bibnamefont{Vion}},
  \bibinfo{author}{\bibfnamefont{D.}~\bibnamefont{Esteve}}, \bibnamefont{and}
  \bibinfo{author}{\bibfnamefont{P.}~\bibnamefont{Bertet}},
  \textit{Hybrid Quantum Circuit with Superconducting Qubit Coupled to a Spin Ensemble},
  \bibinfo{journal}{Phys. Rev. Lett.} \textbf{\bibinfo{volume}{107}},
  \bibinfo{pages}{220501} (\bibinfo{year}{2011})
  
\bibitem[{\citenamefont{Ams\"uss \textit{et~al.}}(2011)\citenamefont{Ams\"uss, Koller, N\"obauer, Putz, Rotter, Sandner, Schneider, Schramb\"ock, Steinhauser, Ritsch, Schmiedmayer, Majer}}]{Amsuess11}
  \bibinfo{author}{\bibfnamefont{R.}~\bibnamefont{Ams\"uss}},
  \bibinfo{author}{\bibfnamefont{C.}~\bibnamefont{Koller}},
  \bibinfo{author}{\bibfnamefont{T.}~\bibnamefont{N\"obauer}},
  \bibinfo{author}{\bibfnamefont{S.}~\bibnamefont{Putz}},
  \bibinfo{author}{\bibfnamefont{S.}~\bibnamefont{Rotter}},
  \bibinfo{author}{\bibfnamefont{K.}~\bibnamefont{Sandner}},
  \bibinfo{author}{\bibfnamefont{S.}~\bibnamefont{Schneider}},
  \bibinfo{author}{\bibfnamefont{M.}~\bibnamefont{Schramb\"ock}},
  \bibinfo{author}{\bibfnamefont{G.}~\bibnamefont{Steinhauser}},
  \bibinfo{author}{\bibfnamefont{H.}~\bibnamefont{Ritsch}},
  \bibinfo{author}{\bibfnamefont{J.}~\bibnamefont{Schmiedmayer}}, \bibnamefont{and}
  \bibinfo{author}{\bibfnamefont{J.}~\bibnamefont{Majer}},
  \textit{Cavity QED with Magnetically Coupled Collective Spin States},
  \bibinfo{journal}{Phys. Rev. Lett.} \textbf{\bibinfo{volume}{107}},
  \bibinfo{pages}{060502} (\bibinfo{year}{2011})
   
\bibitem[{\citenamefont{Probst \textit{et~al.}}(2013)\citenamefont{Probst, Rotzinger, W\"unsch, Jung, Jerger, Siegel, Ustinov, Bushev}}]{Probst13}
  \bibinfo{author}{\bibfnamefont{S.}~\bibnamefont{Probst}},
  \bibinfo{author}{\bibfnamefont{H.}~\bibnamefont{Rotzinger}},
  \bibinfo{author}{\bibfnamefont{S.}~\bibnamefont{W\"unsch}},
  \bibinfo{author}{\bibfnamefont{P.}~\bibnamefont{Jung}},
  \bibinfo{author}{\bibfnamefont{M.}~\bibnamefont{Jerger}},
  \bibinfo{author}{\bibfnamefont{M.}~\bibnamefont{Siegel}},
  \bibinfo{author}{\bibfnamefont{A.~V.}~\bibnamefont{Ustinov}}, \bibnamefont{and}
  \bibinfo{author}{\bibfnamefont{P.}~\bibnamefont{Bushev}},
  \textit{Anisotropic Rare-Earth Spin Ensemble Strongly Coupled to a Superconducting Resonator},
  \bibinfo{journal}{Phys. Rev. Lett.} \textbf{\bibinfo{volume}{110}},
  \bibinfo{pages}{157001} (\bibinfo{year}{2013})
	
\bibitem[{\citenamefont{Grezes \textit{et~al.}}(2014)\citenamefont{Grezes, Julsgaard, Kubo, Stern, Umeda, Isoya, Sumiya, Abe, Onoda, Ohshima, Jacques, Esteve, M{\o}lmer, Bertet}}]{Grezes14}
  \bibinfo{author}{\bibfnamefont{C.}~\bibnamefont{Grezes}},
  \bibinfo{author}{\bibfnamefont{B.}~\bibnamefont{Julsgaard}},
	\bibinfo{author}{\bibfnamefont{Y.}~\bibnamefont{Kubo}},
	\bibinfo{author}{\bibfnamefont{M.}~\bibnamefont{Stern}},
  \bibinfo{author}{\bibfnamefont{T.}~\bibnamefont{Umeda}},
  \bibinfo{author}{\bibfnamefont{J.}~\bibnamefont{Isoya}},
  \bibinfo{author}{\bibfnamefont{H.}~\bibnamefont{Sumiya}},
  \bibinfo{author}{\bibfnamefont{H.}~\bibnamefont{Abe}},
  \bibinfo{author}{\bibfnamefont{S.}~\bibnamefont{Onoda}},
  \bibinfo{author}{\bibfnamefont{T.}~\bibnamefont{Ohshima}},
  \bibinfo{author}{\bibfnamefont{V.}~\bibnamefont{Jacques}},
  \bibinfo{author}{\bibfnamefont{D.}~\bibnamefont{Esteve}},
	\bibinfo{author}{\bibfnamefont{K.}~\bibnamefont{M{\o}lmer}},\bibnamefont{and}
  \bibinfo{author}{\bibfnamefont{P.}~\bibnamefont{Bertet}},
  \textit{Multimode Storage and Retrieval of Microwave Fields in a Spin Ensemble},
  \bibinfo{journal}{Phys. Rev. X} \textbf{\bibinfo{volume}{4}},
  \bibinfo{pages}{021049} (\bibinfo{year}{2014})
	
\bibitem[{\citenamefont{Verd\'u \textit{et~al.}}(2009)\citenamefont{Verd\'u, Zoubi, Koller, Majer, Ritsch, Schmiedmayer}}]{Verdu09}
  \bibinfo{author}{\bibfnamefont{J.}~\bibnamefont{Verd\'u}},
  \bibinfo{author}{\bibfnamefont{H.}~\bibnamefont{Zoubi}},
  \bibinfo{author}{\bibfnamefont{C.}~\bibnamefont{Koller}},
  \bibinfo{author}{\bibfnamefont{J.}~\bibnamefont{Majer}},
  \bibinfo{author}{\bibfnamefont{H.}~\bibnamefont{Ritsch}}, \bibnamefont{and}
  \bibinfo{author}{\bibfnamefont{J.}~\bibnamefont{Schmiedmayer}},
  \textit{Strong Magnetic Coupling of an Ultracold Gas to a Superconducting Waveguide Cavity},
  \bibinfo{journal}{Phys. Rev. Lett.} \textbf{\bibinfo{volume}{103}},
  \bibinfo{pages}{043603} (\bibinfo{year}{2009})
  
\bibitem[{\citenamefont{Henschel \textit{et~al.}}(2010)\citenamefont{Henschel, Majer, Schmiedmayer, Ritsch}}]{Henschel10}
  \bibinfo{author}{\bibfnamefont{K.}~\bibnamefont{Henschel}},
  \bibinfo{author}{\bibfnamefont{J.}~\bibnamefont{Majer}},
  \bibinfo{author}{\bibfnamefont{J.}~\bibnamefont{Schmiedmayer}}, \bibnamefont{and}
  \bibinfo{author}{\bibfnamefont{H.}~\bibnamefont{Ritsch}},
  \textit{Cavity QED with an ultracold ensemble on a chip: Prospects for strong magnetic coupling at finite temperatures},
  \bibinfo{journal}{Phys. Rev. A} \textbf{\bibinfo{volume}{82}},
  \bibinfo{pages}{033810} (\bibinfo{year}{2010})
 
\bibitem[{\citenamefont{Bernon \textit{et~al.}}(2013)\citenamefont{Bernon, Hattermann, Bothner, Knufinke, Weiss, Jessen, Cano, Kemmler, Kleiner, Koelle, Fort\'agh}}]{Bernon13}
  \bibinfo{author}{\bibfnamefont{S.}~\bibnamefont{Bernon}},
  \bibinfo{author}{\bibfnamefont{H.}~\bibnamefont{Hattermann}},
  \bibinfo{author}{\bibfnamefont{D.}~\bibnamefont{Bothner}},
  \bibinfo{author}{\bibfnamefont{M.}~\bibnamefont{Knufinke}},
  \bibinfo{author}{\bibfnamefont{P.}~\bibnamefont{Weiss}},
  \bibinfo{author}{\bibfnamefont{F.}~\bibnamefont{Jessen}},
  \bibinfo{author}{\bibfnamefont{D.}~\bibnamefont{Cano}},
  \bibinfo{author}{\bibfnamefont{M.}~\bibnamefont{Kemmler}},
	\bibinfo{author}{\bibfnamefont{R.}~\bibnamefont{Kleiner}},
  \bibinfo{author}{\bibfnamefont{D.}~\bibnamefont{Koelle}}, \bibnamefont{and}
  \bibinfo{author}{\bibfnamefont{J.}~\bibnamefont{Fort\'agh}},
  \textit{Manipulation and coherence of ultra-cold atoms on a superconducting atom chip},
  \bibinfo{journal}{Nature Commun.} \textbf{\bibinfo{volume}{4}},
  \bibinfo{pages}{2380} (\bibinfo{year}{2013})

\bibitem[{\citenamefont{Jenkins \textit{et~al.}}(2013)\citenamefont{Jenkins, H\"ummer, Mart\'{\i}nez-P\'erez, Garc\'{\i}a-Ripoll, Zueco, Luis}}]{Jenkins13}
  \bibinfo{author}{\bibfnamefont{M.}~\bibnamefont{Jenkins}},
  \bibinfo{author}{\bibfnamefont{T.}~\bibnamefont{H\"ummer}},
  \bibinfo{author}{\bibfnamefont{M.~J.}~\bibnamefont{Mart\'{\i}nez-P\'erez}},
  \bibinfo{author}{\bibfnamefont{J.}~\bibnamefont{Garc\'{\i}a-Ripoll}},
  \bibinfo{author}{\bibfnamefont{D.}~\bibnamefont{Zueco}}, \bibnamefont{and}
  \bibinfo{author}{\bibfnamefont{F.}~\bibnamefont{Luis}},
  \textit{Coupling single-molecule magnets to quantum circuits},
  \bibinfo{journal}{New J. Phys.} \textbf{\bibinfo{volume}{15}},
  \bibinfo{pages}{095007} (\bibinfo{year}{2013})
	
\bibitem[{\citenamefont{Schuster \textit{et~al.}}(2010)\citenamefont{Schuster, Fragner, Dykman, Lyon, Schoelkopf}}]{Schuster10}
  \bibinfo{author}{\bibfnamefont{D.~I.}~\bibnamefont{Schuster}},
  \bibinfo{author}{\bibfnamefont{A.}~\bibnamefont{Fragner}},
  \bibinfo{author}{\bibfnamefont{M.~I.}~\bibnamefont{Dykman}},
  \bibinfo{author}{\bibfnamefont{S.~A.}~\bibnamefont{Lyon}}, \bibnamefont{and}
  \bibinfo{author}{\bibfnamefont{R.~J.}~\bibnamefont{Schoelkopf}},
  \textit{Proposal for Manipulating and Detecting Spin and Orbital States of Trapped Electrons on Helium Using Cavity Quantum Electrodynamics},
  \bibinfo{journal}{Phys. Rev. Lett.} \textbf{\bibinfo{volume}{105}},
  \bibinfo{pages}{040503} (\bibinfo{year}{2010})	

\bibitem[{\citenamefont{Bushev et~al.}(2011)\citenamefont{Bushev, Bothner,
  Nagel, Kemmler, Konovalenko, Loerincz, Ilin, Siegel, Koelle, Kleiner, Schmidt-Kaler}}]{Bushev11}
  \bibinfo{author}{\bibfnamefont{P.}~\bibnamefont{Bushev}},
  \bibinfo{author}{\bibfnamefont{D.}~\bibnamefont{Bothner}},
  \bibinfo{author}{\bibfnamefont{J.}~\bibnamefont{Nagel}},
  \bibinfo{author}{\bibfnamefont{M.}~\bibnamefont{Kemmler}},
  \bibinfo{author}{\bibfnamefont{K.~B.} \bibnamefont{Konovalenko}},
  \bibinfo{author}{\bibfnamefont{A.}~\bibnamefont{Loerincz}},
  \bibinfo{author}{\bibfnamefont{K.}~\bibnamefont{Ilin}},
  \bibinfo{author}{\bibfnamefont{M.}~\bibnamefont{Siegel}},
  \bibinfo{author}{\bibfnamefont{D.}~\bibnamefont{Koelle}}, 
  \bibinfo{author}{\bibfnamefont{R.}~\bibnamefont{Kleiner}}, \bibnamefont{and}
  \bibinfo{author}{\bibfnamefont{F.}~\bibnamefont{Schmidt-Kaler}},
  \textit{Trapped electron coupled to superconducting devices},
  \bibinfo{journal}{Eur. Phys. J. D} \textbf{\bibinfo{volume}{63}},
  \bibinfo{pages}{9} (\bibinfo{year}{2011})

\bibitem[{\citenamefont{Daniilidis \textit{et~al.}}(2013)\citenamefont{Daniilidis, Gorman, Tian, H\"affner}}]{Daniilidis13}
  \bibinfo{author}{\bibfnamefont{N.}~\bibnamefont{Daniilidis}},
  \bibinfo{author}{\bibfnamefont{D.~J.}~\bibnamefont{Gorman}},
  \bibinfo{author}{\bibfnamefont{L.}~\bibnamefont{Tian}}, \bibnamefont{and}
  \bibinfo{author}{\bibfnamefont{H.}~\bibnamefont{H\"affner}},
  \textit{Quantum information processing with trapped electrons and superconducting electronics},
  \bibinfo{journal}{New J. Phys.} \textbf{\bibinfo{volume}{15}},
  \bibinfo{pages}{073017} (\bibinfo{year}{2013})

\bibitem[{\citenamefont{Bushev \textit{et~al.}}(2011)\citenamefont{Bushev, Feofanov, Rotzinger, Protopopov, Cole, Wilson, Fischer, Lukashenko, Ustinov}}]{Bushev11a}
  \bibinfo{author}{\bibfnamefont{P.}~\bibnamefont{Bushev}},
  \bibinfo{author}{\bibfnamefont{A.~K.}~\bibnamefont{Feofanov}},
  \bibinfo{author}{\bibfnamefont{H.}~\bibnamefont{Rotzinger}},
  \bibinfo{author}{\bibfnamefont{I.}~\bibnamefont{Protopopov}},
  \bibinfo{author}{\bibfnamefont{J.~H.}~\bibnamefont{Cole}},
  \bibinfo{author}{\bibfnamefont{C.~M.}~\bibnamefont{Wilson}},
  \bibinfo{author}{\bibfnamefont{G.}~\bibnamefont{Fischer}},
  \bibinfo{author}{\bibfnamefont{A.}~\bibnamefont{Lukashenko}}, \bibnamefont{and}
  \bibinfo{author}{\bibfnamefont{A.~V.}~\bibnamefont{Ustinov}},
  \textit{Ultralow-power spectroscopy of a rare-earth spin ensemble using a superconducting resonator},
  \bibinfo{journal}{Phys. Rev. B} \textbf{\bibinfo{volume}{84}},
  \bibinfo{pages}{060501(R)} (\bibinfo{year}{2011})
  
\bibitem[{\citenamefont{Clauss \textit{et~al.}}(2013)\citenamefont{Clauss, Bothner, Koelle, Kleiner, Bogani, Scheffler, Dressel}}]{Clauss13}
  \bibinfo{author}{\bibfnamefont{C.}~\bibnamefont{Clauss}},
  \bibinfo{author}{\bibfnamefont{D.}~\bibnamefont{Bothner}},
  \bibinfo{author}{\bibfnamefont{D.}~\bibnamefont{Koelle}},
  \bibinfo{author}{\bibfnamefont{R.}~\bibnamefont{Kleiner}},
  \bibinfo{author}{\bibfnamefont{L.}~\bibnamefont{Bogani}},
  \bibinfo{author}{\bibfnamefont{M.}~\bibnamefont{Scheffler}}, \bibnamefont{and}
  \bibinfo{author}{\bibfnamefont{M.}~\bibnamefont{Dressel}},
  \textit{Broadband electron spin resonance from 500~MHz to 40~GHz using superconducting coplanar waveguides},
  \bibinfo{journal}{Appl. Phys. Lett.} \textbf{\bibinfo{volume}{102}},
  \bibinfo{pages}{162601} (\bibinfo{year}{2013})
  
\bibitem[{\citenamefont{Malissa \textit{et~al.}}(2013)\citenamefont{Malissa, Schuster, Tyryshkin, Houck, Lyon}}]{Malissa13}
  \bibinfo{author}{\bibfnamefont{H.}~\bibnamefont{Malissa}},
  \bibinfo{author}{\bibfnamefont{D.~I.}~\bibnamefont{Schuster}},
  \bibinfo{author}{\bibfnamefont{A.~M.}~\bibnamefont{Tyryshkin}},
  \bibinfo{author}{\bibfnamefont{A.~A.}~\bibnamefont{Houck}}, \bibnamefont{and}
  \bibinfo{author}{\bibfnamefont{S.~A.}~\bibnamefont{Lyon}},
  \textit{Superconducting coplanar waveguide resonators for low temperature pulsed electron spin resonance},
  \bibinfo{journal}{Rev. Sci. Instrum.} \textbf{\bibinfo{volume}{84}},
  \bibinfo{pages}{025116} (\bibinfo{year}{2013})

\bibitem[{\citenamefont{Ranjan \textit{et~al.}}(2013)\citenamefont{Ranjan, de Lange, Schutjens, Debelhoir, Groen, Szombati, Thoen, Klapwijk, Hanson, DiCarlo}}]{Ranjan13}
  \bibinfo{author}{\bibfnamefont{V.}~\bibnamefont{Ranjan}},
  \bibinfo{author}{\bibfnamefont{G.}~\bibnamefont{de Lange}},
  \bibinfo{author}{\bibfnamefont{R.}~\bibnamefont{Schutjens}},
  \bibinfo{author}{\bibfnamefont{T.}~\bibnamefont{Debelhoir}},
  \bibinfo{author}{\bibfnamefont{J.~P.}~\bibnamefont{Groen}},
  \bibinfo{author}{\bibfnamefont{D.}~\bibnamefont{Szombati}},
  \bibinfo{author}{\bibfnamefont{D.~J.}~\bibnamefont{Thoen}},
	\bibinfo{author}{\bibfnamefont{T.~M.}~\bibnamefont{Klapwijk}},
  \bibinfo{author}{\bibfnamefont{R.}~\bibnamefont{Hanson}}, \bibnamefont{and}
  \bibinfo{author}{\bibfnamefont{L.}~\bibnamefont{DiCarlo}},
  \textit{Probing Dynamics of an Electron-Spin Ensemble via a Superconducting Resonator},
  \bibinfo{journal}{Phys. Rev. Lett.} \textbf{\bibinfo{volume}{110}},
  \bibinfo{pages}{067004} (\bibinfo{year}{2013})

\bibitem[{\citenamefont{Wiemann \textit{et~al.}}(2015)\citenamefont{Wiemann, Simmendinger, Clauss, Bogani, Bothner, Koelle, Kleiner, Dressel, Scheffler}}]{Wiemann15}
  \bibinfo{author}{\bibfnamefont{Y.}~\bibnamefont{Wiemann}},
  \bibinfo{author}{\bibfnamefont{J.}~\bibnamefont{Simmendinger}},
  \bibinfo{author}{\bibfnamefont{C.}~\bibnamefont{Clauss}},
  \bibinfo{author}{\bibfnamefont{L.}~\bibnamefont{Bogani}},
  \bibinfo{author}{\bibfnamefont{D.}~\bibnamefont{Bothner}},
  \bibinfo{author}{\bibfnamefont{D.}~\bibnamefont{Koelle}},
  \bibinfo{author}{\bibfnamefont{R.}~\bibnamefont{Kleiner}},
  \bibinfo{author}{\bibfnamefont{M.}~\bibnamefont{Dressel}}, \bibnamefont{and}
  \bibinfo{author}{\bibfnamefont{M.}~\bibnamefont{Scheffler}},
  \textit{Observing electron spin resonance between 0.1 and 67~GHz at temperatures between 50~mK and 300~K using broadband metallic coplanar waveguides},
  \bibinfo{journal}{Appl. Phys. Lett.} \textbf{\bibinfo{volume}{106}},
  \bibinfo{pages}{193505} (\bibinfo{year}{2015})

\bibitem[{\citenamefont{Scheffler \textit{et~al.}}(2013)\citenamefont{Scheffler, Schlegel, Clauss, Hafner, Fella, Dressel, Jourdan, Sichelschmidt, Krellner, Geibel, Steglich}}]{Scheffler13}
  \bibinfo{author}{\bibfnamefont{M.}~\bibnamefont{Scheffler}},
  \bibinfo{author}{\bibfnamefont{K.}~\bibnamefont{Schlegel}},
  \bibinfo{author}{\bibfnamefont{C.}~\bibnamefont{Clauss}},
  \bibinfo{author}{\bibfnamefont{D.}~\bibnamefont{Hafner}},
  \bibinfo{author}{\bibfnamefont{C.}~\bibnamefont{Fella}},
  \bibinfo{author}{\bibfnamefont{M.}~\bibnamefont{Dressel}},
  \bibinfo{author}{\bibfnamefont{M.}~\bibnamefont{Jourdan}},
  \bibinfo{author}{\bibfnamefont{J.}~\bibnamefont{Sichelschmidt}},
  \bibinfo{author}{\bibfnamefont{C.}~\bibnamefont{Krellner}},
  \bibinfo{author}{\bibfnamefont{C.}~\bibnamefont{Geibel}}, \bibnamefont{and}
  \bibinfo{author}{\bibfnamefont{F.}~\bibnamefont{Steglich}},
  \textit{Microwave spectroscopy on heavy-fermion systems: Probing the dynamics of charges and magnetic moments},
  \bibinfo{journal}{Phys. Status Solidi B} \textbf{\bibinfo{volume}{250}},
  \bibinfo{pages}{439} (\bibinfo{year}{2013})
	
\bibitem[{\citenamefont{Hafner \textit{et~al.}}(2014)\citenamefont{Hafner, Dressel, Scheffler}}]{Hafner14}
  \bibinfo{author}{\bibfnamefont{D.}~\bibnamefont{Hafner}},
  \bibinfo{author}{\bibfnamefont{M.}~\bibnamefont{Dressel}}, \bibnamefont{and}
  \bibinfo{author}{\bibfnamefont{M.}~\bibnamefont{Scheffler}},
  \textit{Surface-resistance measurements using superconducting stripline resonators},
  \bibinfo{journal}{Rev. Sci. Instrum.} \textbf{\bibinfo{volume}{85}},
  \bibinfo{pages}{014702} (\bibinfo{year}{2014})
		
\bibitem[{\citenamefont{Gittleman and Rosenblum}(1968)}]{Gittleman68}
  \bibinfo{author}{\bibfnamefont{J.~I.}~\bibnamefont{Gittleman}} \bibnamefont{and}
  \bibinfo{author}{\bibfnamefont{B.}~\bibnamefont{Rosenblum}},
  \textit{The Pinning Potential and High-Frequency Studies of Type-II Superconductors},
  \bibinfo{journal}{J. Appl. Phys.} \textbf{\bibinfo{volume}{39}},
  \bibinfo{pages}{2617} (\bibinfo{year}{1968})

\bibitem[{\citenamefont{Brandt}(1991)}]{Brandt91}
  \bibinfo{author}{\bibfnamefont{E.~H.} \bibnamefont{Brandt}},
  \textit{Penetration of Magnetic ac Fields into Type-II Superconductors},
  \bibinfo{journal}{Phys. Rev. Lett.} \textbf{\bibinfo{volume}{67}},
  \bibinfo{pages}{2219} (\bibinfo{year}{1991})

\bibitem[{\citenamefont{Coffey and Clem}(1991)}]{Coffey91}
  \bibinfo{author}{\bibfnamefont{M.~W.} \bibnamefont{Coffey}} \bibnamefont{and}
  \bibinfo{author}{\bibfnamefont{J.~R.} \bibnamefont{Clem}},
  \textit{Unified Theory of Effects of Vortex Pinning and Flux Creep upon the rf Surface Impedance of Type-II Superconductors},
  \bibinfo{journal}{Phys. Rev. Lett.} \textbf{\bibinfo{volume}{82}},
  \bibinfo{pages}{386} (\bibinfo{year}{1991})

\bibitem[{\citenamefont{Pompeo and Silva}(2008)}]{Pompeo08}
  \bibinfo{author}{\bibfnamefont{N.} \bibnamefont{Pompeo}} \bibnamefont{and}
  \bibinfo{author}{\bibfnamefont{E.} \bibnamefont{Silva}},
  \textit{Reliable determination of vortex parameters from measurements of the microwave complex resistivity},
  \bibinfo{journal}{Phys. Rev. B} \textbf{\bibinfo{volume}{78}},
  \bibinfo{pages}{094503} (\bibinfo{year}{2008})
			
\bibitem[{\citenamefont{Song \textit{et~al.}}(2009)\citenamefont{Song,
  Heitmann, DeFeo, Yu, McDermott, Neeley, Martinis, and Plourde}}]{Song09a}
  \bibinfo{author}{\bibfnamefont{C.}~\bibnamefont{Song}},
  \bibinfo{author}{\bibfnamefont{T.~W.} \bibnamefont{Heitmann}},
  \bibinfo{author}{\bibfnamefont{M.~P.} \bibnamefont{DeFeo}},
  \bibinfo{author}{\bibfnamefont{K.}~\bibnamefont{Yu}},
  \bibinfo{author}{\bibfnamefont{R.}~\bibnamefont{McDermott}},
  \bibinfo{author}{\bibfnamefont{M.}~\bibnamefont{Neeley}},
  \bibinfo{author}{\bibfnamefont{J.~M.} \bibnamefont{Martinis}}, \bibnamefont{and}
	\bibinfo{author}{\bibfnamefont{B.~L.~T.}~\bibnamefont{Plourde}},
	\textit{Microwave response of vortices in superconducting thin films of Re and Al},
	\bibinfo{journal}{Phys. Rev. B} \textbf{\bibinfo{volume}{79}},
	\bibinfo{pages}{174512} (\bibinfo{year}{2009}{\natexlab{a}})
				
\bibitem[{\citenamefont{Bothner \textit{et~al.}}(2011)\citenamefont{Bothner,
  Gaber, Kemmler, Koelle, and Kleiner}}]{Bothner11}
  \bibinfo{author}{\bibfnamefont{D.}~\bibnamefont{Bothner}},
  \bibinfo{author}{\bibfnamefont{T.}~\bibnamefont{Gaber}},
  \bibinfo{author}{\bibfnamefont{M.}~\bibnamefont{Kemmler}},
  \bibinfo{author}{\bibfnamefont{D.}~\bibnamefont{Koelle}},
  \bibnamefont{and} \bibinfo{author}{\bibfnamefont{R.} \bibnamefont{Kleiner}},
  \textit{Improving the performance of superconducting microwave resonators in magnetic fields},
  \bibinfo{journal}{Appl. Phys. Lett.} \textbf{\bibinfo{volume}{98}},
  \bibinfo{pages}{102504} (\bibinfo{year}{2011})
	
\bibitem[{\citenamefont{Andrews and Mathew}(2012)}]{Andrews12}
\bibinfo{author}{\bibfnamefont{J.} \bibnamefont{Andrews}} \bibnamefont{and}
  \bibinfo{author}{\bibfnamefont{V.} \bibnamefont{Mathew}},
  \textit{Magnetic field induced properties of type II superconducting microstrip resonators},
  \bibinfo{journal}{Supercond. Sci. Technol.} \textbf{\bibinfo{volume}{25}},
  \bibinfo{pages}{025004} (\bibinfo{year}{2012})	
	
\bibitem[{\citenamefont{Healey \textit{et~al.}}(2008)\citenamefont{Healey, Lindstr\"om, Colclough, Muirhead, Tzalenchuk}}]{Healey08}
\bibinfo{author}{\bibfnamefont{J.~E.}~\bibnamefont{Healey}},
  \bibinfo{author}{\bibfnamefont{T.}~\bibnamefont{Lindstr\"om}},
  \bibinfo{author}{\bibfnamefont{M.~S.}~\bibnamefont{Colclough}},
  \bibinfo{author}{\bibfnamefont{C.~M.}~\bibnamefont{Muirhead}}, \bibnamefont{and}
	\bibinfo{author}{\bibfnamefont{A.~Ya.} \bibnamefont{Tzalenchuk}},
	\textit{Magnetic field tuning of coplanar microwave resonators},
  \bibinfo{journal}{Appl. Phys. Lett.} \textbf{\bibinfo{volume}{93}},
  \bibinfo{pages}{043513} (\bibinfo{year}{2008})
	
\bibitem[{\citenamefont{Song \textit{et~al.}}(2009)\citenamefont{Song, DeFeo, Yu, Plourde}}]{Song09}
\bibinfo{author}{\bibfnamefont{C.}~\bibnamefont{Song}},
  \bibinfo{author}{\bibfnamefont{M.~P.} \bibnamefont{DeFeo}},
  \bibinfo{author}{\bibfnamefont{K.}~\bibnamefont{Yu}}, \bibnamefont{and}
  \bibinfo{author}{\bibfnamefont{B.~L.~T.}~\bibnamefont{Plourde}},
  \textit{Reducing microwave loss in superconducting resonators due to trapped vortices},
  \bibinfo{journal}{Appl. Phys. Lett.} \textbf{\bibinfo{volume}{95}},
  \bibinfo{pages}{232501} (\bibinfo{year}{2009})
	
\bibitem[{\citenamefont{Groll \textit{et~al.}}(2010)\citenamefont{Groll, Gurevich, Chiorescu}}]{Groll10}
  \bibinfo{author}{\bibfnamefont{N.}~\bibnamefont{Groll}},
  \bibinfo{author}{\bibfnamefont{A.}~\bibnamefont{Gurevich}}, \bibnamefont{and}
  \bibinfo{author}{\bibfnamefont{I.}~\bibnamefont{Chiorescu}},
  \textit{Measurement of the nonlinear Meissner effect in superconducting Nb films using a resonant microwave cavity: A probe of unconventional pairing symmetries},
  \bibinfo{journal}{Phys. Rev. B} \textbf{\bibinfo{volume}{81}},
  \bibinfo{pages}{020504(R)} (\bibinfo{year}{2010})
	
\bibitem[{\citenamefont{de Graaf \textit{et~al.}}(2012)\citenamefont{de Graaf, Danilov, Adamyan, Bauch, Kubatkin}}]{deGraaf12}
  \bibinfo{author}{\bibfnamefont{S.~E.}~\bibnamefont{de Graaf}},
  \bibinfo{author}{\bibfnamefont{A.~V.}~\bibnamefont{Danilov}},
  \bibinfo{author}{\bibfnamefont{A.}~\bibnamefont{Adamyan}},
  \bibinfo{author}{\bibfnamefont{T.}~\bibnamefont{Bauch}}, \bibnamefont{and}
	\bibinfo{author}{\bibfnamefont{S.~E.} \bibnamefont{Kubatkin}},
	\textit{Magnetic field resilient superconducting fractal resonators for coupling to free spins},
  \bibinfo{journal}{J. Appl. Phys.} \textbf{\bibinfo{volume}{112}},
  \bibinfo{pages}{123905} (\bibinfo{year}{2012})

\bibitem[{\citenamefont{Bothner \textit{et~al.}}(2012)\citenamefont{Bothner, Clauss, Koroknay, Kemmler, Gaber, Jetter, Scheffler, Michler, Dressel, Koelle, and Kleiner}}]{Bothner12}
  \bibinfo{author}{\bibfnamefont{D.}~\bibnamefont{Bothner}},
  \bibinfo{author}{\bibfnamefont{C.}~\bibnamefont{Clauss}},
  \bibinfo{author}{\bibfnamefont{E.}~\bibnamefont{Koroknay}},
  \bibinfo{author}{\bibfnamefont{M.}~\bibnamefont{Kemmler}},
  \bibinfo{author}{\bibfnamefont{T.}~\bibnamefont{Gaber}},
  \bibinfo{author}{\bibfnamefont{M.}~\bibnamefont{Jetter}},
  \bibinfo{author}{\bibfnamefont{M.}~\bibnamefont{Scheffler}},
  \bibinfo{author}{\bibfnamefont{P.}~\bibnamefont{Michler}},
  \bibinfo{author}{\bibfnamefont{M.}~\bibnamefont{Dressel}},
  \bibinfo{author}{\bibfnamefont{D.}~\bibnamefont{Koelle}}, \bibnamefont{and}
	\bibinfo{author}{\bibfnamefont{R.} \bibnamefont{Kleiner}},
	\textit{Reducing vortex losses in superconducting microwave resonators with microsphere patterned antidot arrays},
  \bibinfo{journal}{Appl. Phys. Lett.} \textbf{\bibinfo{volume}{100}},
  \bibinfo{pages}{012601} (\bibinfo{year}{2012})
	
\bibitem[{\citenamefont{Bothner \textit{et~al.}}(2012)\citenamefont{Bothner, Gaber, Kemmler, Koelle, and Kleiner, W\"unsch, Siegel}}]{Bothner12a}
  \bibinfo{author}{\bibfnamefont{D.}~\bibnamefont{Bothner}},
  \bibinfo{author}{\bibfnamefont{T.}~\bibnamefont{Gaber}},
	\bibinfo{author}{\bibfnamefont{M.}~\bibnamefont{Kemmler}},
  \bibinfo{author}{\bibfnamefont{D.}~\bibnamefont{Koelle}},
  \bibinfo{author}{\bibfnamefont{R.}~\bibnamefont{Kleiner}},
  \bibinfo{author}{\bibfnamefont{S.}~\bibnamefont{W\"unsch}}, \bibnamefont{and}
	\bibinfo{author}{\bibfnamefont{M.}~\bibnamefont{Siegel}},
	\textit{Magnetic hysteresis effects in superconducting coplanar microwave resonators},
  \bibinfo{journal}{Phys. Rev. B} \textbf{\bibinfo{volume}{86}},
  \bibinfo{pages}{014517} (\bibinfo{year}{2012})

\bibitem[{\citenamefont{Ghirri \textit{et~al.}}(2015)\citenamefont{Ghirri, Bonizzoni, Gerace, Sanna, Cassinese, Affronte}}]{Ghirri15}
  \bibinfo{author}{\bibfnamefont{A.}~\bibnamefont{Ghirri}},
  \bibinfo{author}{\bibfnamefont{C.}~\bibnamefont{Bonizzoni}},
	\bibinfo{author}{\bibfnamefont{D.}~\bibnamefont{Gerace}},
  \bibinfo{author}{\bibfnamefont{S.}~\bibnamefont{Sanna}},
  \bibinfo{author}{\bibfnamefont{A.}~\bibnamefont{Cassinese}}, \bibnamefont{and}
	\bibinfo{author}{\bibfnamefont{M.}~\bibnamefont{Affronte}},
	\textit{YBa2Cu3O7 microwave resonators for strong collective coupling with spin ensembles},
  \bibinfo{journal}{Appl. Phys. Lett.} \textbf{\bibinfo{volume}{106}},
  \bibinfo{pages}{184101} (\bibinfo{year}{2015})
  
\bibitem[{\citenamefont{Samkharadze \textit{et~al.}}(2015)\citenamefont{Samkharadze, Bruno, Scarlino, Zheng, DiVicenzo, DiCarlo, Vandersypen}}]{Samkharadze15}
  \bibinfo{author}{\bibfnamefont{N.}~\bibnamefont{Samkharadze}},
  \bibinfo{author}{\bibfnamefont{A.}~\bibnamefont{Bruno}},
  \bibinfo{author}{\bibfnamefont{G.}~\bibnamefont{Zheng}},
	\bibinfo{author}{\bibfnamefont{P.}~\bibnamefont{Scarlino}},
  \bibinfo{author}{\bibfnamefont{D.~P.}~\bibnamefont{DiVincenzo}},
  \bibinfo{author}{\bibfnamefont{L.}~\bibnamefont{DiCarlo}}, \bibnamefont{and}
	\bibinfo{author}{\bibfnamefont{L.~M.~K.}~\bibnamefont{Vandersypen}},
	\textit{High kinetic inductance superconducting nanowire resonators for circuit QED in a magnetic field}
  \bibinfo{journal}{Phys. Rev. Applied} \textbf{\bibinfo{volume}{5}},
  \bibinfo{pages}{044004} (\bibinfo{year}{2016})

\bibitem[{\citenamefont{Bothner \textit{et~al.}}(2013)\citenamefont{Bothner, Knufinke, Hattermann, W\"olbing, Ferdinand, Weiss, Bernon, Fort\'agh, Koelle, Kleiner}}]{Bothner13}
  \bibinfo{author}{\bibfnamefont{D.}~\bibnamefont{Bothner}},
  \bibinfo{author}{\bibfnamefont{M.}~\bibnamefont{Knufinke}},
	\bibinfo{author}{\bibfnamefont{H.}~\bibnamefont{Hattermann}},
  \bibinfo{author}{\bibfnamefont{R.}~\bibnamefont{W\"olbing}},
  \bibinfo{author}{\bibfnamefont{B.}~\bibnamefont{Ferdinand}},
	\bibinfo{author}{\bibfnamefont{P.}~\bibnamefont{Weiss}},
  \bibinfo{author}{\bibfnamefont{S.}~\bibnamefont{Bernon}},
  \bibinfo{author}{\bibfnamefont{J.}~\bibnamefont{Fort\'agh}},
  \bibinfo{author}{\bibfnamefont{D.}~\bibnamefont{Koelle}}, \bibnamefont{and}
	\bibinfo{author}{\bibfnamefont{R.}~\bibnamefont{Kleiner}},
	\textit{Inductively coupled superconducting half wavelength resonators as persistent current traps for ultracold atoms},
  \bibinfo{journal}{New J. Phys.} \textbf{\bibinfo{volume}{15}},
  \bibinfo{pages}{093024} (\bibinfo{year}{2013})
	
\bibitem[{\citenamefont{Cano \textit{et~al.}}(2008)\citenamefont{Cano, Kasch, Hattermann, Kleiner, Zimmermann, Koelle, Fort\'agh}}]{Cano08}
  \bibinfo{author}{\bibfnamefont{D.}~\bibnamefont{Cano}},
  \bibinfo{author}{\bibfnamefont{B.}~\bibnamefont{Kasch}},
	\bibinfo{author}{\bibfnamefont{H.}~\bibnamefont{Hattermann}},
  \bibinfo{author}{\bibfnamefont{R.}~\bibnamefont{Kleiner}},
  \bibinfo{author}{\bibfnamefont{C.}~\bibnamefont{Zimmermann}},
  \bibinfo{author}{\bibfnamefont{D.}~\bibnamefont{Koelle}}, \bibnamefont{and}
	\bibinfo{author}{\bibfnamefont{J.}~\bibnamefont{Fort\'agh}},
	\textit{Meissner Effect in Superconducting Microtraps},
  \bibinfo{journal}{Phys. Rev. Lett.} \textbf{\bibinfo{volume}{101}},
  \bibinfo{pages}{183006} (\bibinfo{year}{2008})
  
\bibitem[{\citenamefont{Supplementary Information}(2017)}]{Supp16}
  \bibinfo{author}{\bibfnamefont{}\bibnamefont{}}
  \bibinfo{author}{\bibfnamefont{}\bibnamefont{}}
  \textit{Supplementary Information}
  \bibinfo{journal}{}\textbf{\bibinfo{volume}{}}
  \bibinfo{pages}{}(\bibinfo{year}{2017})
  
\bibitem[{\citenamefont{Bosman \textit{et~al.}}(2015)\citenamefont{Bosman, Singh, Bruno, Steele}}]{Bosman15}
  \bibinfo{author}{\bibfnamefont{S.~J.}~\bibnamefont{Bosman}},
  \bibinfo{author}{\bibfnamefont{V.}~\bibnamefont{Singh}},
  \bibinfo{author}{\bibfnamefont{A.}~\bibnamefont{Bruno}}, \bibnamefont{and}
	\bibinfo{author}{\bibfnamefont{G.~A.}~\bibnamefont{Steele}},
	\textit{Broadband architecture for galvanically accessible superconducting microwave resonators},
  \bibinfo{journal}{Appl. Phys. Lett.} \textbf{\bibinfo{volume}{107}},
  \bibinfo{pages}{192602} (\bibinfo{year}{2015})
 
\bibitem[{\citenamefont{G\"oppl \textit{et~al.}}(2008)\citenamefont{G\"oppl, Fragner, Baur, Bianchetti, Filipp, Fink, Leek, Puebla, Steffen, Wallraff}}]{Goeppl08}
  \bibinfo{author}{\bibfnamefont{M.}~\bibnamefont{G\"oppl}},
  \bibinfo{author}{\bibfnamefont{A.}~\bibnamefont{Fragner}},
	\bibinfo{author}{\bibfnamefont{M.}~\bibnamefont{Baur}},
  \bibinfo{author}{\bibfnamefont{R.}~\bibnamefont{Bianchetti}},
  \bibinfo{author}{\bibfnamefont{S.}~\bibnamefont{Filipp}},
	\bibinfo{author}{\bibfnamefont{J.~M.}~\bibnamefont{Fink}},
  \bibinfo{author}{\bibfnamefont{P.~J.}~\bibnamefont{Leek}},
  \bibinfo{author}{\bibfnamefont{G.}~\bibnamefont{Puebla}},
  \bibinfo{author}{\bibfnamefont{L.}~\bibnamefont{Steffen}}, \bibnamefont{and}
	\bibinfo{author}{\bibfnamefont{A.}~\bibnamefont{Wallraff}},
	\textit{Coplanar waveguide resonators for circuit quantum electrodynamics},
  \bibinfo{journal}{J. Appl. Phys.} \textbf{\bibinfo{volume}{104}},
  \bibinfo{pages}{113904} (\bibinfo{year}{2008})
  
\bibitem[{\citenamefont{Khapaev \textit{et~al.}}(2003)\citenamefont{Khapaev, Kupriyanov, Goldobin, Siegel}}]{Khapaev03}
  \bibinfo{author}{\bibfnamefont{M.~M.}~\bibnamefont{Khapaev}},
  \bibinfo{author}{\bibfnamefont{M.~Y.}~\bibnamefont{Kupriyanov}},
	\bibinfo{author}{\bibfnamefont{E.}~\bibnamefont{Goldobin}}, \bibnamefont{and}
	\bibinfo{author}{\bibfnamefont{M.}~\bibnamefont{Siegel}},
	\textit{Current distribution simulation for superconducting multi-layered structures},
  \bibinfo{journal}{Supercond. Sci. Technol.} \textbf{\bibinfo{volume}{16}},
  \bibinfo{pages}{24} (\bibinfo{year}{2003})
  
\bibitem[{\citenamefont{Clem}(2013)}]{Clem13}
  \bibinfo{author}{\bibfnamefont{John~R.}~\bibnamefont{Clem}},
  \textit{Inductances and attenuation constant for a thin-film superconducting coplanar waveguide resonator},
  \bibinfo{journal}{J. Appl. Phys.} \textbf{\bibinfo{volume}{113}},
  \bibinfo{pages}{013910} (\bibinfo{year}{2013})
  
\bibitem[{\citenamefont{Javaheri Rahim \textit{et~al.}}(2016)\citenamefont{Javaheri Rahim, Lehleiter, Bothner, Krellner, Koelle, Kleiner, Dressel, Scheffler}}]{Javaheri16}
  \bibinfo{author}{\bibfnamefont{M.}~\bibnamefont{Javaheri~Rahim}},
  \bibinfo{author}{\bibfnamefont{T.}~\bibnamefont{Lehleiter}},
  \bibinfo{author}{\bibfnamefont{D.}~\bibnamefont{Bothner}},
  \bibinfo{author}{\bibfnamefont{C.}~\bibnamefont{Krellner}},
  \bibinfo{author}{\bibfnamefont{D.}~\bibnamefont{Koelle}},
  \bibinfo{author}{\bibfnamefont{R.}~\bibnamefont{Kleiner}},
	\bibinfo{author}{\bibfnamefont{M.}~\bibnamefont{Dressel}}, \bibnamefont{and}
	\bibinfo{author}{\bibfnamefont{M.}~\bibnamefont{Scheffler}},
	\textit{Metallic coplanar resonators optimized for low-temperature measurements},
  \bibinfo{journal}{Journal of Physics D: Applied Physics} \textbf{\bibinfo{volume}{49}},
  \bibinfo{pages}{395501} (\bibinfo{year}{2016})
  
\bibitem[{\citenamefont{Norris}(1970)}]{Norris70}
  \bibinfo{author}{\bibfnamefont{W.~T.}~\bibnamefont{Norris}},
  \textit{Calculation of hysteresis losses in hard superconductors carrying ac: isolated conductors and edges of thin sheets},
  \bibinfo{journal}{J. Phys. D: Appl. Phys.} \textbf{\bibinfo{volume}{3}},
  \bibinfo{pages}{489} (\bibinfo{year}{1970})

\bibitem[{\citenamefont{Brandt and Indenbom}(1993)}]{Brandt93}
  \bibinfo{author}{\bibfnamefont{E.~H.} \bibnamefont{Brandt}} \bibnamefont{and}
  \bibinfo{author}{\bibfnamefont{M.}~\bibnamefont{Indenbom}},
  \textit{Type-II-superconducting strip with current in a perpendicular magnetic field},
  \bibinfo{journal}{Phys. Rev. B} \textbf{\bibinfo{volume}{48}},
  \bibinfo{pages}{12893} (\bibinfo{year}{1993})
  
\bibitem[{\citenamefont{Wenner \textit{et~al.}}(2011)\citenamefont{Wenner, Neeley, Bialczak, Lenander, Lucero, O'Connell, Sank, Wang, Weides, Cleland, Martinis}}]{Wenner11}
\bibinfo{author}{\bibfnamefont{J.}~\bibnamefont{Wenner}},
  \bibinfo{author}{\bibfnamefont{M.}~\bibnamefont{Neeley}},
  \bibinfo{author}{\bibfnamefont{R.~C.}~\bibnamefont{Bialczak}},
	\bibinfo{author}{\bibfnamefont{M.}~\bibnamefont{Lenander}},
  \bibinfo{author}{\bibfnamefont{E.}~\bibnamefont{Lucero}},
  \bibinfo{author}{\bibfnamefont{A.~D.}~\bibnamefont{O'Connell}},
	\bibinfo{author}{\bibfnamefont{D.}~\bibnamefont{Sank}},
  \bibinfo{author}{\bibfnamefont{H.}~\bibnamefont{Wang}},
  \bibinfo{author}{\bibfnamefont{M.}~\bibnamefont{Weides}},
  \bibinfo{author}{\bibfnamefont{A.~N.}~\bibnamefont{Cleland}}, \bibnamefont{and}
	\bibinfo{author}{\bibfnamefont{J.~M.}~\bibnamefont{Martinis}},
	\textit{Wirebond crosstalk and cavity modes in large chip mounts for superconducting qubits},
  \bibinfo{journal}{Supercond. Sci. Technol.} \textbf{\bibinfo{volume}{24}},
  \bibinfo{pages}{065001} (\bibinfo{year}{2011})  
\end{thebibliography}
\end{document}


\title{Supplementary Material for: Improving superconducting resonators in magnetic fields by reduced field-focussing and engineered flux screening}

\author{D.~Bothner}
\author{D.~Wiedmaier}
\author{B.~Ferdinand}
\author{R.~Kleiner}
\author{D.~Koelle}

\maketitle

\section*{Spectra of resonators without normal-conducting ground-plane extensions}

\begin{figure}[h]
\centering {\includegraphics[scale=0.6]{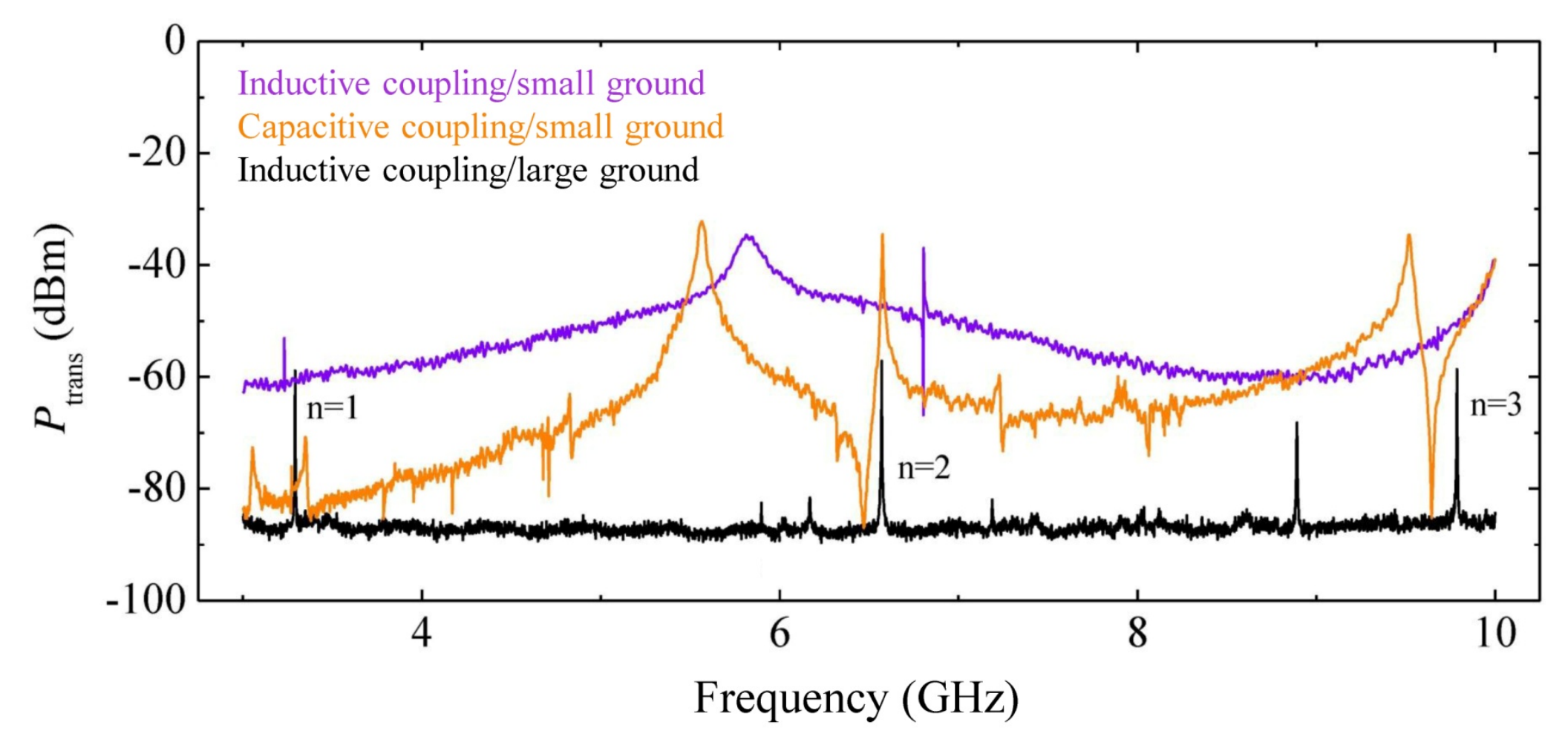}}
\caption{Transmission spectra of three different microwave resonators between 3~GHz and 10~GHz. The resonator with large ground-planes has $S=50\,\mu$m and $W=25\,\mu$m. The two resonators with small superconducting ground conductors have $S=50\,\mu$m, $W=25\,\mu$m and $G=50\,\mu$m. The applied power was $P_\mathrm{app}=-20\,$dBm, measurement temperature was $T=4.2\,$K and applied magnetic field was $B=0$. The mode number of the resonator with full ground-planes is indicated by $n$.}
\label{fig:FigureSupp1}
\end{figure}

In the main text, we describe measurements on coplanar microwave resonators which have been fabricated by a combination of superconducting and normal-conducting films.
These resonators consist of a superconducting center conductor with width $S$ and superconducting ground conductors on both sides of the center conductor, but the superconducting ground conductors have only a width of $G=50\,\mu$m.
The rest of the ground-planes is made of a normal-conducting film, cf. Fig.~1~(c), (d) of the main text.
The width of the gap between center conductor and ground planes is $W$.
Here, we show and discuss the transmission spectra of resonators with $G=50\,\mu$m but without the normal-conducting parts.
As a consequence of not having the normal-conducting parts, the ground conductors are only connected to the sample box at the transition from the microwave connectors to the chip.
Figure~\ref{fig:FigureSupp1} shows in direct comparison the transmitted microwave power for three different resonator configurations.
The black line with a constant background transmission of $\sim -88\,$dBm shows the spectrum of an inductively coupled resonator with $S=50\,\mu$m and superconducting ground-planes to the edge of the chip (IC-\textbf{N}-50, cf. main text).
At the chip edge, the ground-planes are everywhere around the chip connected to the sample box walls by means of silver paste.
The spectrum shows three very pronounced transmission peaks around 3.3, 6.6 and 9.9~GHz, which correspond to the first three modes of the $\lambda/2$ resonator.
Except for some small spurious resonances far away from the resonator peaks, the transmission is completely suppressed.
A very different transmission is observed for the two resonators with narrow ground-planes.
There is a large transmission over almost the whole frequency range; for certain frequency ranges the transmission is even orders of magnitude higher than the transmission in resonance of the full-ground sample.
This observation reveals that for small ground-planes the microwave finds a way around the waveguide resonator through the box.
Most probably the microwave fields occupy the whole box volume, which is very unfavorable for any kind of experiment, where mode selectivity or spatial field confinement is desired or necessary.
From the large and broad transmission peaks between $5$ and $6\,$GHz and another increase around 10~GHz, we come to the conclusion that low quality factor box resonances are probably responsible for this high broadband transmission. 
These box resonances, however, are not just a background but they are coupled to the coplanar microwave structure.
They depend on its geometry and coupling type, which is revealed by the differences in frequency and width of the peak for the two different resonators.
It is important to note, that the geometrical difference between the two coupling types is very small compared to the chip or the box dimensions.
What is important is the change of boundary conditions due to the different coupling types, and this change of boundary conditions is significant and has an influence on the transmission over the complete frequency range.
The difference in transmission between the inductively coupled and the capacitively coupled resonator is systematic and qualitatively reproducible for identical boxes, mountings and resonator layouts, so it is not related to different qualities of the transition from the SMA connectors to the chip or other non-systematic variations between different chips.
For other resonator layouts, other cross-sectional parameters, and other box geometries, however, also the background transmission changes significantly (data not shown here).
The presence of the box modes and the fact that they are coupled to the coplanar microwave structures, however, has another consequence.
Around the resonance frequencies of the full-ground coplanar microwave resonator (indicated with $n=1, 2, 3$ in the figure), resonance signatures also appear in the transmission spectra of the small-ground samples.
But they are significantly distorted in magnitude and frequency.
This distortion reveals that we do not see bare coplanar waveguide resonances anymore; the coplanar resonators are reactively and dissipatively loaded by the box walls.
Thus, a fair comparison between these loaded modes and the bare coplanar waveguide modes with respect to the impact of magnetic fields is not possible.
How the coplanar waveguide resonances are loaded and how the resonance frequencies and quality factors are shifted, however, depends on mode shape, box shape, coupling type and probably more, i.e., on many parameters of the complete system.
In conclusion we state, that without a proper grounding of the waveguide ground conductors, the transmission becomes sensitive to many parameters of the system and the resonances can not be attributed to the coplanar waveguide structure alone anymore.
When using normal-conducting ground-plane extensions, however, the background transmission is suppressed again to the level of the resonator with full ground and the resonance frequencies are back at their designed positions, independent of coupling type, mode number etc.
The implementation of our approaches without normal-conducting elements thus requires also a careful consideration of sample holder design, chip mounting and resonator layout.